\documentclass[conference]{IEEEtran}
\IEEEoverridecommandlockouts
\usepackage{fancyhdr} 

\usepackage[utf8]{inputenc}
\usepackage[T1]{fontenc}

\usepackage{subfig}
\usepackage{float}
\usepackage[noadjust]{cite}
\usepackage{graphicx}
\usepackage[export]{adjustbox}

\usepackage{microtype}

\usepackage{balance}
\usepackage[normalem]{ulem}
\usepackage{enumitem}

\usepackage{tabularx}
\usepackage{booktabs}
\usepackage{multirow}

\usepackage[noadjust]{cite}

\usepackage{amsmath,amsfonts}

\usepackage[noend]{algpseudocode}
\usepackage{algorithm}

\usepackage{balance}

\newcommand{\clustername}{Prometheus}

\author{%
    \IEEEauthorblockN{%
        Bartłomiej Przybylski\IEEEauthorrefmark{1},
        Maciej Pawlik\IEEEauthorrefmark{2}\IEEEauthorrefmark{3},
        Paweł Żuk\IEEEauthorrefmark{1},\\
        Bartłomiej Łagosz\IEEEauthorrefmark{2},
        Maciej Malawski\IEEEauthorrefmark{4}\IEEEauthorrefmark{2}, and 
        Krzysztof Rzadca\IEEEauthorrefmark{1}%
        \thanks{Corresponding author: B. Przybylski}
    }%
    \IEEEauthorblockA{%
        \IEEEauthorrefmark{1}\emph{University of Warsaw, Institute of Informatics}, Warsaw, Poland, \\
        \IEEEauthorrefmark{2}\emph{AGH University of Science and Technology, Institute of Computer Science}, Krakow, Poland,\\
        \IEEEauthorrefmark{3}\emph{Academic Computer Centre Cyfronet AGH}, Krakow, Poland, \\
        \IEEEauthorrefmark{4}\emph{Sano Centre for Computational Medicine}, Krakow, Poland \\
        E-mail: bap@mimuw.edu.pl, mapawlik@agh.edu.pl, p.zuk@mimuw.edu.pl,\\blagosz@student.agh.edu.pl, malawski@agh.edu.pl, krzadca@mimuw.edu.pl
    }
}

\begin{document}

\title{Using Unused: Non-Invasive Dynamic FaaS Infrastructure with HPC-Whisk}

\maketitle

\thispagestyle{fancy}
\lhead{}
\rhead{}
\chead{}
\lfoot{\footnotesize{
Preprint of the paper accepted at SC22, Dallas, Texas, USA, 2022.\\
\textcopyright{} 2022 IEEE. Personal use of this material is permitted. Permission from IEEE must be obtained for all other uses, in any current or future media, including  reprinting/republishing this material for advertising or promotional purposes, creating new collective works, for resale or redistribution to servers or lists, or reuse of any copyrighted component of this work in other works.}}
\rfoot{}
\cfoot{}
\renewcommand{\headrulewidth}{0pt}
\renewcommand{\footrulewidth}{0pt}

\begin{abstract}
Modern HPC workload managers and their careful tuning contribute to the high utilization of HPC clusters. However, due to inevitable uncertainty it is impossible to completely avoid node idleness. Although such idle slots are usually too short for any HPC job, they are too long to ignore them. Function-as-a-Service (FaaS) paradigm promisingly fills this gap, and can be a good match, as typical FaaS functions last seconds, not hours. Here we show how to build a FaaS infrastructure on idle nodes in an HPC cluster in such a way that it does not affect the performance of the HPC jobs significantly. We dynamically adapt to a changing set of idle physical machines, by integrating open-source software Slurm and OpenWhisk. 

We designed and implemented a prototype solution that allowed us to cover up to 90\% of the idle time slots on a 50k-core cluster that runs production workloads.
\end{abstract}


\section{Introduction}
\label{sec:introduction}

Function-as-a-Service (FaaS) is a serverless cloud computing paradigm in which clients deploy individual functions and providers schedule the function executions~\cite{Serverless2019}. Although high-performance computing (HPC), high-throughput (HTC) and FaaS are all used in scientific computing, their workload characteristics are substantially different.
In HPC, a job is assigned a requested number of nodes (and perhaps other resources) for a given period of time (usually hours) and has to manage these resources on its own. The provider has no job-level responsibility; instead, they maximize the overall utilization of the cluster expressed as a fraction of nodes allocated to jobs over time. Similarly, HTC opportunistically assigns resources to very long jobs that consist of loosely-coupled tasks; HTC jobs need to be managed by the client on their own.
Classic cloud computing approaches --- IaaS and PaaS --- also fall within this broad category. The cloud provider furnishes VMs or runs Kubernetes containers following the customer's requests on their shape: the number of requested VMs/containers, their requested number of cores, and the amount of memory. The VMs or containers run until the customer decides to stop them (hours to days).

In contrast, FaaS is fine-grained: a FaaS end-user issues individual invocations of functions, rather than complete jobs; and the FaaS provider allocates resources for each invocation. Thus, from the programmer's perspective, FaaS is often easier to use and maintain than monolithic applications. Additionally, from the provider's perspective, FaaS  operates on a much finer time-scale than a standard HPC job.
While we have no aggregate data on FaaS \emph{scientific} workloads, we might get some intuition from the commercial Azure Functions workload~\cite{Shahrad2020}: 50\% of the functions completed \emph{in less than 3s}, and 90\% in less than 1 minute. This sharply contrasts with HPC jobs: on \clustername, a large production HPC cluster operated by Academic Computer Centre Cyfronet AGH, a median job \emph{declares} a runtime of 60 minutes, and 95\% of jobs declare at least 15 minutes (Fig.~\ref{fig:joblog}). In IaaS clouds, time scales are similar to HPC: in the Azure VM trace \cite{AzureVMTrace} VMs running for at least \emph{3 days} account for more than 95\% of the total core-hours.

Although modern workload managers for HPC and HTC (e.g., Slurm~\cite{yoo2003slurm}) implement efficient algorithms for resource management and scheduling, even with backfilling \cite{backfill-feitelson,estimation-backfill,characterization-backfill} it is nearly impossible to use all the nodes at all times.
First, jobs' requested number of nodes varies from  1 to a significant share of the whole cluster. Second, it is difficult to accurately predict the state of the environment even in the nearest future. 
This stems from (inherently difficult to predict) users' actions like submitting or canceling jobs, changing their priorities, inaccurate estimations of job duration~\cite{fan2017trade,tang2010analyzing} (slack in Fig.~\ref{fig:joblog}), 
and external factors (e.g., node failures).
The queuing system on \clustername{} is typical: tuned over years, its daily average utilization is usually over 99\% --- but never 100\%.

To summarize, (1) despite the well-tuned HPC schedulers, supercomputers still have idle nodes; and (2) in contrast to hours-long HPC and IaaS jobs, the FaaS workload consists of many short-lived function invocations, which, in principle, should act as ``sand'' filling up the gaps between HPC jobs' ``rocks''. We study the following questions:

\begin{enumerate}[label={Q\arabic*},leftmargin=*]
\item Can we effectively use these unused resources of an HPC cluster for FaaS?
\item How much computing capacity can we gain with this approach?
\end{enumerate}

\begin{figure*}[t]
    \centering
    \subfloat[CDF of the number of idle nodes. E.g., 20\% of time there were at most 2 idle nodes (bottom left); and 80\% of time, there were at most 13 idle nodes. X-axis limited to the 99\%ile (67 nodes).]{
        \includegraphics[height=0.18\textwidth,clip,valign=t]{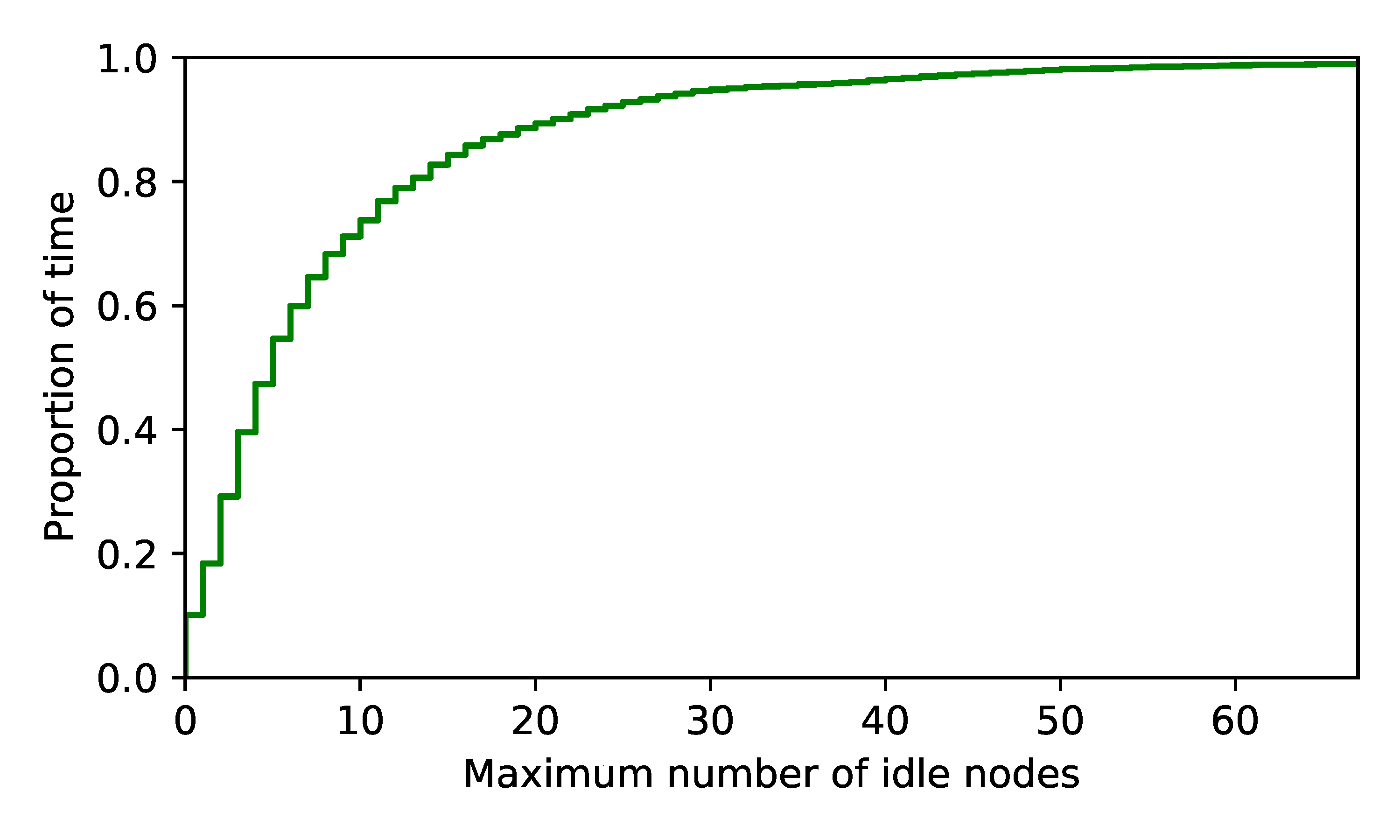}
        \label{fig:Prometheus-IdleNodes-CDF}
    }\hspace*{0.5em}%
    \subfloat[CDF of the length of idleness periods]{
        \includegraphics[height=0.18\textwidth,clip,valign=t]{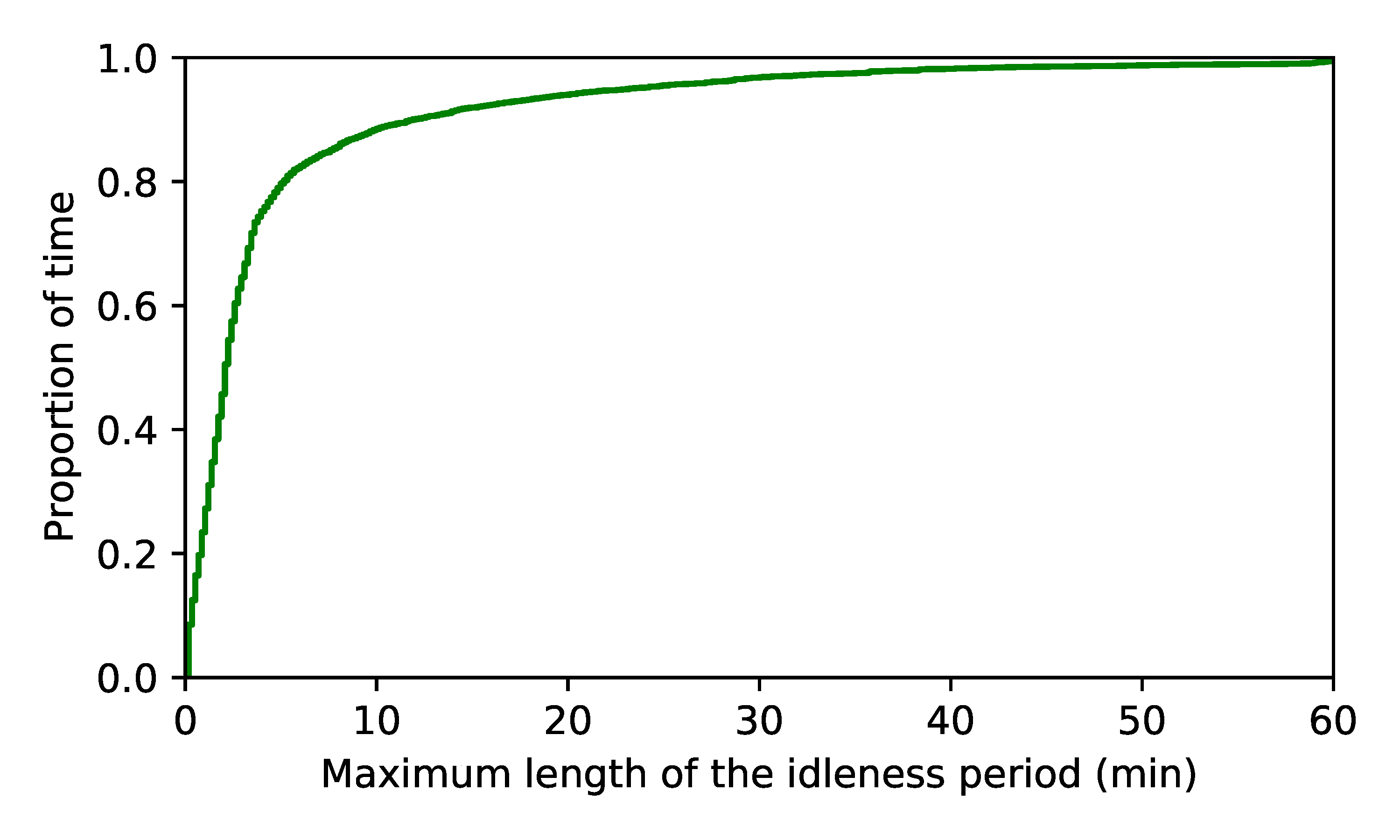}
        \label{fig:HPC-IP}
    }\hspace*{0.5em}%
    \subfloat[Time series of the number of idle nodes over time. Note rapid changes, with short bursts of up to 150 idle nodes.]{
        \includegraphics[height=0.18\textwidth,clip,valign=t]{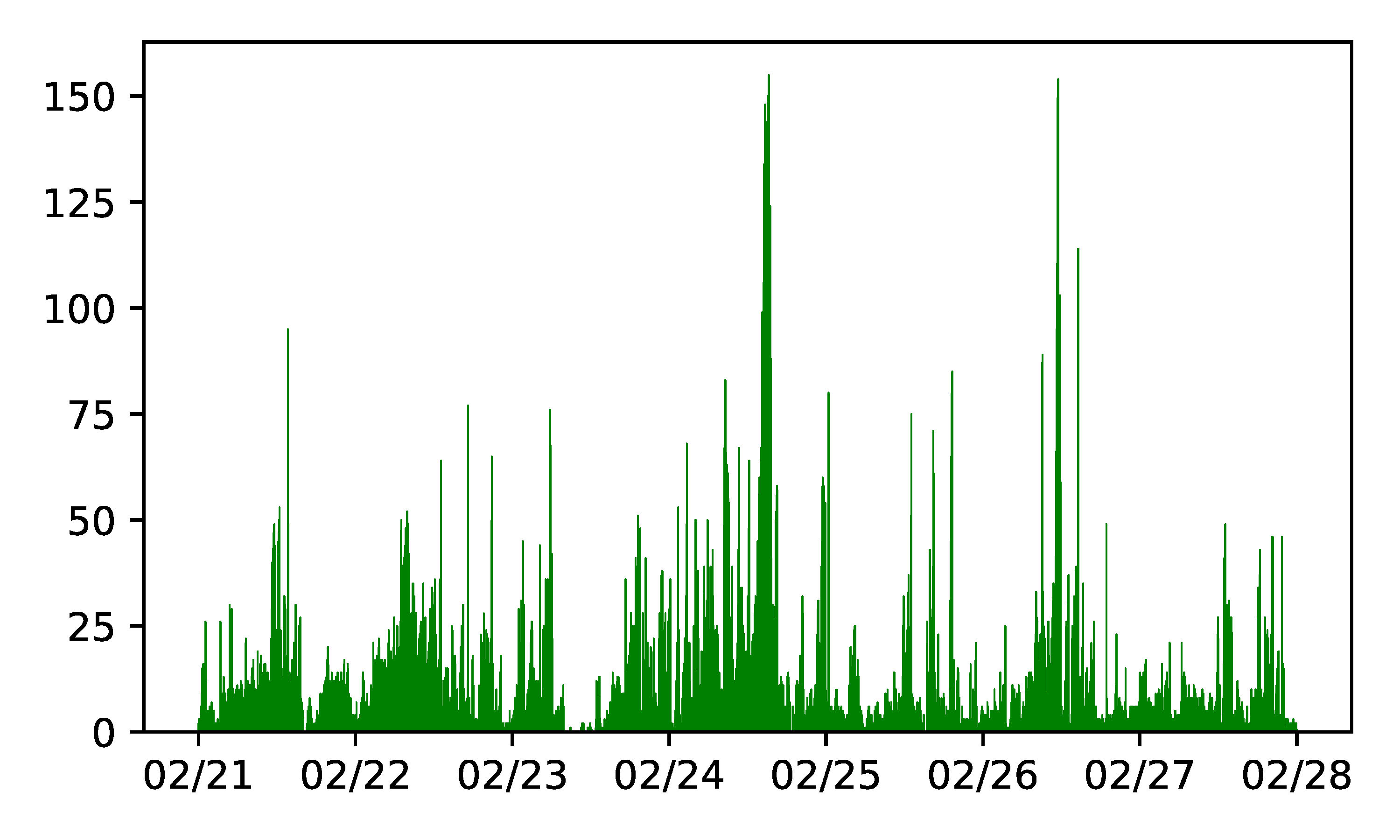}
        \label{fig:Prometheus-IdleNodes-T}
    }
      
    \caption{Analysis of \clustername{} workload for 21-27 Feb. 2022. The charts are based on regularly-logged Slurm node statuses and exclude nodes allocated to commercial reservations.}\label{fig:clusterlog}
\end{figure*}

\begin{figure*}
    \centering
    \begin{minipage}{0.32\linewidth}
    \centering
    \includegraphics[width=0.8\textwidth]{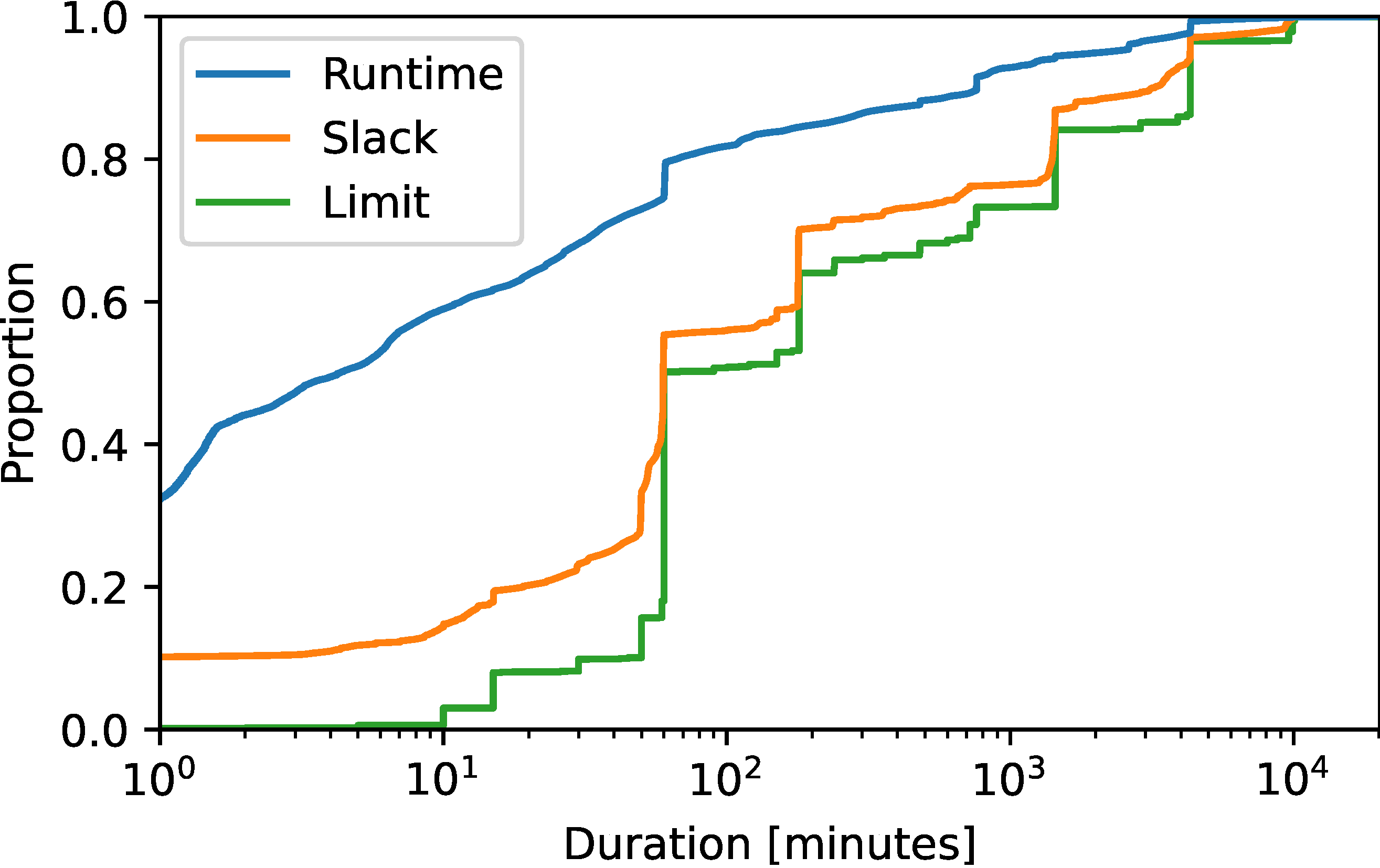}
    \caption{CDFs of: user-declared time limits for HPC jobs (green); their runtimes (blue); the differences between the limit and the runtime, or the slack (orange). In the monitored period, 74k non-commercial jobs were completed.}
    \label{fig:joblog}
    \end{minipage}
    \hfill
    \begin{minipage}{0.65\linewidth}
    \includegraphics[width=\textwidth]{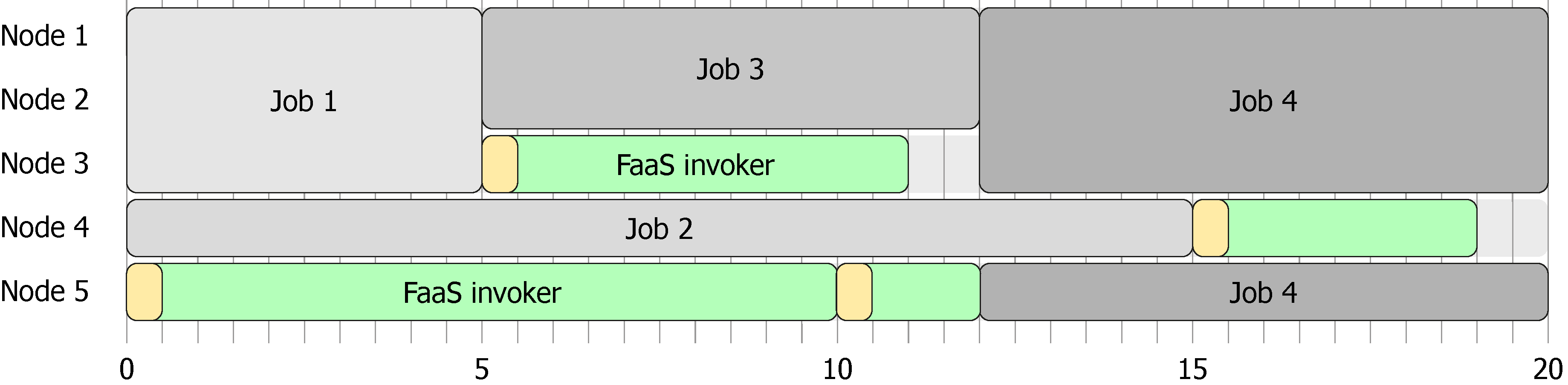}
    \caption{An assignment of $4$ HPC jobs (gray) to $5$ nodes that minimize the maximum completion time, but still leaves substantial idle time (e.g. node 5 is idle until time moment 12). In yellow/green we present potential short-term, single-node FaaS invoker jobs that could fill the gaps.}
    \label{fig:HPC-Ex1}
    \end{minipage}
\end{figure*}

Consider the following example. A 5-node HPC cluster executes $4$ jobs. The 1st job reserves $3$ nodes for $5$ minutes; the 2nd job --- $1$ node for $13$ minutes; the 3rd job --- $2$ nodes for $7$ minutes; and the 4th job --- $4$ nodes for $8$ minutes. 
Fig.~\ref{fig:HPC-Ex1} shows a schedule with a minimal length.
Even assuming perfect estimates and no state variability, the average number of idle nodes is $1.2$. Our approach partially fills the gaps with short, single-node pilot jobs acting as FaaS invokers and processing function invocations (Fig.~\ref{fig:HPC-Ex1}). 
A FaaS invoker needs some time to initialize (e.g., to register with the controller, marked in yellow), and then processes incoming function call requests (green).
With FaaS invoker jobs of lengths 2, 4, 6 and 10 minutes (see Sec.~\ref{subsec:lengths}), submitted as low-priority and backfilled, ready invokers cover 83\% of previously idle slots, \emph{without any impact on the prime HPC workload}.

To quantify this opportunity on a production system, we analyzed \clustername{} job execution logs. The main partition of this cluster consists of 2,239 nodes equipped with two 12-core Intel Xeon E5-2680v3 processors and 128 GB of RAM.
We monitored node usage for a week (February 21st-27th, 2022: we confirmed with the administrators that this week was typical). \clustername{} processes both scientific and commercial jobs. Scientific jobs are managed by Slurm, while commercial customers reserve blocks of nodes for long periods, which are managed separately: no scientific job can be executed on an idle, yet reserved node. Thus, we excluded all the commercial nodes from our analysis, regardless of their status. 

The cluster was highly-loaded: over nodes not reserved for commercial jobs, the average utilization \emph{exceeded 99\%}. Yet, on the average, $9.23$ nodes were idle at any moment (with the 25th percentile of 2 and a median of 5, Fig.~\ref{fig:Prometheus-IdleNodes-CDF}).
This equals to a considerable total idle \emph{surface} of over 37,000 core-hours. % 9.23 * 24 * 7 * 24
However, for roughly $10.11\%$ of time, ca. $17$ out of $168$ hours, not a single node was available. The longest unavailability period was $1.55$ hour. Fig.~\ref{fig:Prometheus-IdleNodes-T} shows the distribution over time.

From the perspective of any single node, the idleness periods revealed a very dynamic setting. \emph{The median duration of node idleness was just 2 minutes, and the 75th percentile was about 4 minutes} (see Fig.~\ref{fig:HPC-IP}). The distribution had a long tail, as the mean was slightly over 5 minutes and 5\% of idleness periods were longer than 23 minutes. Thus, while there is a significant idle \emph{surface}, it is composed of many short periods, making it impossible to use by most HPC jobs (95\% of which declare runtime of at least 15 minutes, Fig.~\ref{fig:joblog}), or e.g. for provisioning  VMs. 
These periods are also much shorter (thus more difficult to use) than AWS Spot instances~\cite{10.1145/3126908.3126953} (which are sometimes used for dynamic FaaS infrastructures).

While too short for HPC, the 2--4 minute periods are, in principle, sufficient for typical, seconds-long FaaS invocations. 
However, the brevity of these periods puts a significant strain on the FaaS infrastructure:
before processing invocations, a node must register into the FaaS cluster;
after the node is requested for the prime, HPC workload, 
it must stop accepting new invocations, off-load its queue and de-register.

The principal contribution of this paper is the design and engineering of \emph{HPC-Whisk}, a FaaS layer seamlessly running on a standard HPC cluster. To the best of our knowledge, HPC-Whisk is the first FaaS system optimized to harvest idle HPC nodes. HPC-Whisk fulfills the three primary design goals:

\begin{enumerate}[leftmargin=*]
    \item \emph{HPC-Whisk is minimally invasive on the HPC infrastructure}.
    We do not modify the system software running the cluster and we do not claim resources beyond the already-existing idle periods.
    HPC-Whisk invokers are submitted to Slurm as low-priority, preemptible jobs with either fixed, or variable lengths. 
    As HPC-Whisk uses existing Slurm configuration options (preemption, variable job lengths), it requires little effort from HPC system administrators and no trust in additional software modules --- which should help in practical applicability of our solution.
    \item \emph{HPC-Whisk is convenient for the FaaS users}. 
    Rather than proposing custom APIs or requesting a specific method to submit the workload, we extend OpenWhisk, a standard FaaS middleware.
    We thus provide the same interface to the end-user and to the developer of FaaS functions; and a familiar system software to the FaaS DevOps engineer. The primary engineering challenge here is that HPC-Whisk must swiftly react to dynamic changes in the available infrastructure. A new invoker must be quickly added to the system. And, when Slurm decides to terminate an invoker to prepare resources for a prime HPC workload, this invoker needs to off-load its requests and de-register.
    \item \emph{HPC-Whisk is efficient} at converting these best-effort, per-node idle periods into a productive FaaS infrastructure.
\end{enumerate}

Answering questions Q1-Q2, our experiments show that:

\begin{enumerate}[label={A\arabic*},leftmargin=*]
    \item With fixed-sized jobs, HPC-Whisk turns 87\% %10.39/10.66 * 0.9 
    of the idle surface into healthy OpenWhisk workers capable of executing invocations. Our implementation efficiently handles nodes' dynamic arrivals and departures: compared with a posteriori simulation upper-bound, just 5\% % 10.39/10.90
    of the surface is used for such accounting.
    Moreover, our responsiveness experiments show that the HPC-Whisk successfully handles over 95\% of the requests. 
    \item This idle FaaS surface represents significant computational capacity. We run standard computationally-intensive FaaS benchmark on \clustername{}: we observed a consistent, 15\% improvement in response time compared to AWS Lambda.
\end{enumerate}

The paper is organized as follows. Sec.~\ref{sec:background} introduces FaaS architecture and aspects later used in the paper. Sec.~\ref{sec:design} describes the design and key implementation aspects of HPC-Whisk. The following two sections experimentally evaluate HPC-Whisk. Sec.~\ref{sec:calibration} contains preliminary measurements and calibrations of the prototype, notably optimizing the lengths of the pilot jobs. Sec.~\ref{sec:efficiency} measures the performance of HPC-Whisk on the production cluster: how efficient is HPC-Whisk in converting idle periods into ready FaaS workers; how stable is the resulting infrastructure from the perspective of a FaaS user; and how the computational efficiency of a single FaaS worker compares to AWS Lambda. Finally, Sec.~\ref{sec:related-work} discusses related work in (a) using serverless for HPC workloads; (b) scavenging idle HPC resources; (c) using pilot jobs.

\begin{figure*}
    \centering
    \includegraphics[width=0.93\textwidth]{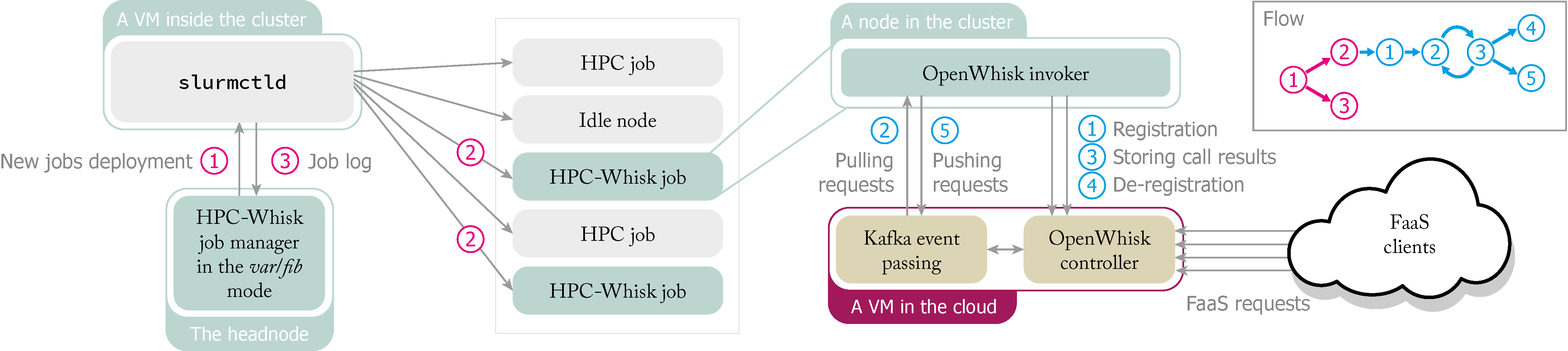}
    \caption{HPC-Whisk architecture. In grey, unmodified HPC cluster components. In green, our additions and modifications. In brown, OpenWhisk components maintained outside of the cluster. Encircled numbers denote the order of event occurrence.}
    \label{fig:slurm_ow_architecture}
\end{figure*}

\section{Background: An Architecture of a Function-as-a-Service (FaaS) System}
\label{sec:background}

This section introduces the Function-as-a-Service aspects we later use in the paper. We refer to~\cite{Serverless2019,hassan_survey_2021} for more details.

Cloud computing covers a wide range of services. In a classic IaaS (Infrastructure-as-a-Service), one creates self-managed virtual machines (VM) for any purpose. Predefined amounts of resources are always reserved for VMs, even if they are not fully utilized. FaaS lies on the other side of the scale. In FaaS, the customer deploys code snippets: stateless functions that are executed in response to internal or external calls. FaaS services scale to zero, i.e. the resources are used only when the function is being executed (or has just finished). This is so, as a container ready to execute a single function can be quickly created and removed by the cloud provider, if needed. 
Commercial FaaS services like Google Functions, Amazon Lambda or Microsoft Azure Functions bill their customers just for the consumed resources. The price is usually based on the number of function calls and their total execution time. Thus, if the customers do not call the function at all, they do not pay for the service, even though the source code of the function is stored by the service provider and can be used at any time. 

Apache OpenWhisk \cite{OpenWhisk} is an open-source platform managing FaaS infrastructure. 
OpenWhisk is actively developed and also offered commercially (IBM Cloud Functions, Adobe I/O Runtime); it is extensively studied (e.g.~\cite{shahrad_architectural_2019,banaei2022etas,9499544}).

A customer uploads stateless functions --- written e.g. in Python or JavaScript --- to OpenWhisk. These functions may be triggered by HTTP requests or other functions. Compared to HPC jobs, functions require very few resources. In fact, a single physical machine is capable of hosting dozens or even hundreds of functions simultaneously. OpenWhisk isolates such functions by maintaining Docker containers that are initialized with requested environments. If there is a high request load on a function, multiple containers for the same function may be maintained, possibly on multiple nodes. On the other hand, if the function has not been called recently, there might be no containers ready to execute it. However, such a container is created (usually in less than 500 milliseconds) when the function is called again.
One of the key roles of OpenWhisk is to manage function containers running on a predefined set of worker nodes, while the load of the deployed functions changes over time.

A standard OpenWhisk setup consists of three types of components: controllers, invokers and containers. Each \emph{controller} is responsible for routing function calls to \emph{invokers} (worker nodes) of its choice, via dedicated Apache Kafka topics. Thus, the controller acts as a load balancer if more than a single invoker is available at the moment. Each invoker manages a single  physical (or virtual)
machine. The resources provided by this machine are used to host multiple \emph{containers} which execute function calls and return the results to the controller. The invoker pulls, on a regular basis, new requests from its individual Apache Kafka topic, and then it executes them one after another (a FIFO strategy is used here). For each request, the invoker itself decides on which container to execute this request. It is possible that no container will be able to execute a given function call. In such a case, a new container may be warmed up, possibly after some unused containers are removed.

As a rule, the controller always tries to pass calls of the same function to the same invoker(s). Usually, the target invoker is determined based on the hashed name of the function. It maximizes the probability that the selected invoker hosts a warmed-up container for such a function, which reduces the time needed to process the request. However, this convenient property has its flaws. In a standard OpenWhisk setup, it is assumed that the set of available invokers does not reduce in time. Thus, any unexpected event, i.e. worker node failure (or termination), may result in no answers to some of the calls. Moreover, when the invoker pulls the requests from its Apache Kafka topic, this decision cannot be easily reversed.

\section{HPC-Whisk: Architecture and Design\\of our Setup}
\label{sec:design}

In this section we describe the main technical contributions of this work, namely the HPC-Whisk, a system we developed on top of OpenWhisk and Slurm to make unused cluster nodes available to FaaS. While designing our setup, we had three main goals in mind, as described in Sec.~\ref{sec:introduction}: (1) minimal invasiveness; (2) convenience for the FaaS users; (3) efficiency. To achieve the first goal, we made our HPC-Whisk jobs act as low-priority and preemptible Slurm jobs (Sec.~\ref{subsec:setup-architecture}). We also made sure that there are always enough HPC-Whisk jobs ready to be allotted on otherwise-idle nodes (Sec.~\ref{subsec:amdq}). The second goal was achieved by running a modified version of the OpenWhisk software on top of our setup, and thus providing the users with known, well-grounded and cluster-independent API interface for FaaS (Sec.~\ref{subsec:setup-singularity}, \ref{sec:ow-dynamic-workers} and \ref{subsec:setup-availability}). Finally, the third goal was achieved by tuning HPC-Whisk jobs so they cover the highest possible share of idleness periods with the lowest possible overhead of warming up (Sec.~\ref{sec:calibration}).

\subsection{Architecture overview}
\label{subsec:setup-architecture}

In our setup, the number and location of OpenWhisk invokers (workers) dynamically change over time --- following the transient nature of idle resources of a busy HPC cluster. 
Thus, as soon as a cluster node becomes available, we try to set up there a new OpenWhisk invoker. Similarly, when a (prime) HPC job demands nodes and Slurm notifies the OpenWhisk invoker about the upcoming termination, we need to clean up the system and re-route all the queued invocations to other workers. 
We achieved these features by (1) modifying OpenWhisk; and (2) supplying job submission scripts for Slurm (and tuning its configuration to run them on a strict best-effort basis). 

Fig.~\ref{fig:slurm_ow_architecture} presents the architecture of our system. 
Our HPC-Whisk job manager (that may work in two modes) submits low-priority, preemptible jobs to the standard \texttt{slurmctld} daemon. The job manager is running on the head node, which provides Slurm interface used for job submission and management. HPC-Whisk jobs require a single physical node, and declare a short (no more than 2 hours) running time compared to standard HPC jobs. Our Slurm configuration guarantees that HPC-Whisk jobs will be placed in otherwise-idle nodes only.
When Slurm starts an HPC-Whisk job, this job starts a local OpenWhisk invoker which immediately registers to the cluster-wide OpenWhisk controller (hosted off-cluster, on a dedicated VM, as it has to run 24/7 in an environment with unconstrained network connectivity), and then executes function calls for between 2 and 90 minutes. On the other hand, when the HPC-Whisk job terminates, it de-registers from the controller. All the unprocessed requests are transferred to other available invokers. To make these state changes possible, our modified OpenWhisk controller maintains a dynamic list of active HPC-Whisk invokers, together with their corresponding Kafka queues. The implementation details, concerning the integration of OpenWhisk and Slurm, are presented in Sec. B-E.

\subsection{Replacing Docker with Singularity}
\label{subsec:setup-singularity}

The goal of HPC-Whisk is to be minimally invasive on the HPC cluster's system software --- and the first resulting system challenge we had to solve was the difference in the containerization methods used in OpenWhisk and in HPC.
OpenWhisk workers maintain and execute functions in Docker containers. Docker containers, while widely adopted in cloud computing, typically require a root-user node daemon --- and maintaining such daemons on cluster nodes would be a nuisance for cluster administrators and can be a source of security concerns.
In contrast, most HPC providers adopted Singularity~\cite{Singulatiry} as the containerization method. Singularity does not require a root-user node daemon, and can be used as a standalone application.
There is work underway for Docker to support rootless execution~\cite{DockerRoot}, but it is not yet widely adopted on HPC systems including \clustername{}. There are other container solutions that are also rootless, e.g. Podman, but they often depend on other system features that are not common for HPC. In the case of Podman, there is a strong dependency on the availability of user-space \texttt{cgroup} support which is not always present, especially on older systems. Thus, we adapted OpenWhisk's node invoker to use Singularity~\cite{damian-msc}.
Note that, from the end user's perspective, the switch from Docker to Singularity is almost transparent. Docker images can be executed by Singularity, except for some of the more advanced network and isolation functionalities.
Also, as soon as~\cite{DockerRoot} becomes available on a cluster, HPC-Whisk can use it.

\subsection{Adding support for dynamic workers}
\label{sec:ow-dynamic-workers}

OpenWhisk support for dynamically-changing set of worker nodes is limited. Once the nodes are up, it is always assumed to be due to a stop failure when any of them stops working. In such a case, the requests placed in the worker-related Kafka topic are not processed at all and, at the end, the timeout error is returned by the controller. Such behaviour is not acceptable in our setup where worker nodes' mean time to ``failure'' (i.e., when the node is requested by a prime HPC job) is in the order of minutes, rather than days. To cope with such frequent failures, we extend OpenWhisk to support both the dynamic appearance and disappearance of invokers.

Slurm notifies a job before terminating it (because of eviction or timeout). In our case, Slurm sends a \texttt{SIGTERM} 3 minutes before the final \texttt{SIGKILL}. We implement a few new mechanisms in OpenWhisk to automatically reconfigure the infrastructure in a few seconds after the initial \texttt{SIGTERM}.

When the HPC-Whisk job receives a \texttt{SIGTERM}, it immediately passes it to the OpenWhisk invoker which performs the following actions. First, it informs the controller that it is not available for new requests anymore. We extended the set of regular messages sent from workers to controllers so the exact status of each worker node is known to the controller continuously. Second, the worker moves all the requests that it had pulled from the Kafka topic, but has not executed yet, from its internal buffer to a global \emph{fast lane} topic. Simultaneously, the controller moves all the unpulled requests from the worker's Kafka topic to the \emph{fast lane} topic. If the worker is executing a function, it interrupts this execution and puts a corresponding request in the \emph{fast lane} topic, too. 
Such interruption might result in an inconsistent state of the client system (e.g. if a function non-atomically modifies the state that is stored externally). We thus allow OpenWhisk client not to use that mechanism --- however, for function calls longer than 3 minutes, it may result in a failed execution and a timeout error returned to the client. 
An additional fault-tolerant layer can be also designed~\cite{sreekanti_fault-tolerance_2020}.

The \emph{fast lane} topic is a global Kafka topic common to all the workers. Before a worker pulls new requests from its individual Kafka topic, it first pulls requests from the \emph{fast lane} topic. This way, requests reissued by a terminating worker are executed with the highest priority.

\subsection{Managing a dynamic queue of HPC-Whisk jobs}
\label{subsec:amdq}

Many HPC clusters, including \clustername{}, are managed by Slurm Workload Manager~\cite{yoo2003slurm}. Slurm uses a centralized service, \texttt{slurmctld} (a controller), to monitor and manage resources. Each of the nodes in the cluster runs a \texttt{slurmd} daemon which executes the decisions of the controller. However, it is \texttt{slurmctld} that is responsible for receiving jobs from the users, scheduling them, releasing them to the cluster nodes, and terminating jobs in response to certain internal events. As executing HPC jobs is the primary purpose of the cluster, our FaaS infrastructure fully integrates with Slurm.

The integration of our HPC-Whisk setup with an existing HPC infrastructure required us to solve two orthogonal problems. First, we did not want HPC-Whisk jobs to worsen the experience of standard HPC jobs. Second, we needed to provide a continuous supply of HPC-Whisk jobs so that all the periods of idleness could be potentially filled.

\paragraph{Priority tiers and preemption}
We configured Slurm so that HPC-Whisk jobs use a partition with \texttt{PriorityTier} of 0, the lowest possible, while other partitions, used for standard HPC jobs, always have the \texttt{PriorityTier} of 1 or more. Slurm never allots a job with a lower priority tier if it would delay any job with a higher priority tier. 
In order not to make future jobs wait for more than usual, we let Slurm evict (preempt) our HPC-Whisk jobs, and release the resources as required.
This behaviour is determined by the \texttt{PreemptMode=CANCEL} parameter of a partition.
As discussed earlier, Slurm gives a 3-minute grace period to a job being evicted --- which delays an HPC job by at most 3 minutes. 
We claim that a 3-minute slowdown is not significant for (usually longer-running) HPC jobs --- however, if needed, this could be further reduced in the Slurm configuration.
As a consequence, HPC-Whisk jobs never significantly dislodge HPC jobs and, in fact, HPC-Whisk jobs are running on nodes that would not be used by any other HPC jobs at the moment.

\paragraph{Supply of OpenWhisk jobs}
A single HPC-Whisk job acts as an OpenWhisk worker. Thus, while executing, it can process many function calls. However, proper choice of the length of such an HPC-Whisk job can significantly influence the system's efficiency.
An HPC-Whisk job needs to initialize before it can start processing. Thus, to maximize the time available for processing and minimize this  overhead, OpenWhisk worker should be running for as long as possible.

As shown in the Introduction, in an HPC cluster the set of idle nodes changes very dynamically. Thus, it is hard to precisely estimate how many HPC-Whisk jobs will be actually executed in a given period of time.
This is so as, in our setup, existing HPC-Whisk jobs are constantly terminated and new ones are allocated. We consider two models of supplying HPC-Whisk jobs to Slurm: \emph{var} and \emph{fib}. For each of the models, we introduce an external job manager that supplies Slurm with HPC-Whisk jobs. In both cases, the job manager is implemented as a shell script application, utilizing the available job management commands, mimicking the standard user interaction with the cluster. The script is Slurm compliant, therefore highly portable, and implements the core functionalities of the chosen job control model.

In the case of the \emph{var} model, we release bags of identical HPC-Whisk jobs with a flexible execution time between 2 minutes (the size of an allocation slot) and 2 hours (backfill window duration) each. By the flexible execution time, we mean that jobs can be elongated or shortened to fit into periods of available resources. We achieve this by specifying the \verb|--time-min| and \verb|--time| job parameters. Slurm determines the exact length of the job during the scheduling process, based on the resources available in the given period and job's minimum and maximum time. We regularly make sure that there are 100 such flexible jobs in the Slurm queue. The queue is replenished with new jobs, if needed, in 15-second intervals.

In the case of the \emph{fib} model, we release jobs with fixed execution times, in the range between 2 and 90 minutes (in Sect.~\ref{subsec:lengths} we show how we chose the exact job lengths). This time scheduler is responsible only for finding an adequate space and time for running jobs with specified, fixed lengths. The higher the execution time, the higher the job's priority within its priority tier. This forces Slurm to use a greedy approach when assigning jobs to long periods of idleness. Regularly, we make sure that there are 10 jobs of each length in the Slurm queue. Again, we replenish the queue with new jobs in 15-second intervals.

In both cases of \emph{var} and \emph{fib} models, the total number of HPC-Whisk jobs in a queue never exceeds 100 at any moment, so the jobs do not introduce a significant load on the Slurm scheduler. This fact is essential as Slurm needs to be able to process the complete job queue within a specific time limit to perform efficient scheduling. It is noteworthy that, although we continuously supply HPC-Whisk jobs to Slurm, we create new jobs only to replace ones that have already started.

One may argue that instead of maintaining a queue of jobs, we could make Slurm create jobs on the fly when it recognizes idle periods after scheduling. However, Slurm doesn't provide such functionality out of the box. Implementing such a solution would significantly alter a scheduler working in a production environment against policies imposed by the HPC site where we developed and tested HPC-Whisk.

\subsection{Dealing with non-availability periods}
\label{subsec:setup-availability}

Our initial 7-day analysis (see Introduction) showed that for 10.11\% of the time no node was idle. It means that for at least the same amount of time, which was around 17 hours, no function could be executed on any potential OpenWhisk invoker in the \clustername{} cluster. In fact, the longest period of full cluster utilization was almost 93 minutes. At the same time, the median and average length of such a period were almost 1 and 3 minutes, respectively.

When there is no worker available, the OpenWhisk controller immediately returns the 503 (Service Unavailable) error in response to any incoming request. For this reason, we propose a client-side wrapper for FaaS calls in the solutions based on our setup, presented as Alg.~\ref{alg:wrapper}. The wrapper simply off-loads FaaS calls to a commercial cloud (e.g. AWS Lambda) for a short duration of time (e.g. 1 minute) when OpenWhisk returns 503. Such an approach removes the risk of starvation of the function calls when no HPC-Whisk jobs are running on a cluster. On the other hand, as a function is short-running, the number of workers rarely changes significantly between submission and invocation. In such a case, however, functions already accepted by OpenWhisk are rerouted to a priority \emph{fast lane} if their current worker stops (see Sec.~\ref{sec:ow-dynamic-workers}).

\begin{algorithm}
    \caption{A wrapper for FaaS function calls with a \clustername{} service of unknown availability.}
    \label{alg:wrapper}
    \begin{algorithmic}
        \State{$\text{Last\_503} \gets \text{datetime("1970-01-01 00:00:00")}$}
        
        \Function{Wrapper}{function, arguments}
            \If{$(\text{datetime.now()} - \text{Last\_503}).\text{seconds()} \leq 60$}
                \State{$r \gets \text{Commercial}.\text{Execute(function, arguments)}$}
            \Else
                \State{$r \gets \text{\clustername{}}.\text{Execute(function, arguments)}$}
                \If{$r.\text{code} = 503$}
                    \State{$\text{Last\_503} \gets \text{datetime.now()}$}
                    \State{\Return \textsc{Wrapper}(function, arguments)}
                \EndIf
            \EndIf
            \State{\Return r}
        \EndFunction
    \end{algorithmic}
\end{algorithm}

\section{Experimental Setup and Calibration}
\label{sec:calibration}

This and the following section experimentally evaluate our prototype.
In this section, we discuss the method (Sec.~\ref{exps:method}) and then perform preliminary measurements and experiments that calibrate the parameters of the system and its simulator (Sec.~\ref{subsec:lengths}). 

\subsection{Method}
\label{exps:method}

All of our experiments are performed on a working cluster. On the one hand, the OpenWhisk controller knows exactly when the invokers start and stop being available, but it does not know anything about idle nodes on the cluster level. On the other hand, Slurm knows everything about jobs it schedules but does not monitor the inner statuses of these jobs. For this reason, we take two different perspectives while monitoring the system during our experiments.

\begin{enumerate}[leftmargin=*]
    \item \emph{OpenWhisk-level.} We analyze the actual performance of our setup based on the second-accurate logs from the OpenWhisk controller and Slurm. By combining these two sources, we are able to determine the status of each HPC-Whisk job up to the second.
    \item \emph{Slurm-level.} We estimate how HPC-Whisk jobs cover node availability periods in the cluster, based on the lists of both idle nodes and HPC-Whisk nodes logged every 10 seconds. The list of idle nodes could not be determined based on second-accurate logs as they only cover the details about busy nodes. At the same time, the set of idle nodes is not necessarily the complement of the set of busy nodes (some nodes may be temporarily unavailable for Slurm due to their maintenance or failure).
\end{enumerate}

For our \emph{Slurm-level} perspective, we repeatedly asked Slurm for a list of available nodes. We aimed to log the state of the cluster every 10 seconds. However, as we used the same approach during the initial analysis (see Sect.~\ref{sec:introduction}), we concluded that it was not possible. The time Slurm required to process our log requests between 21st and 27th Feb. 2022 varied from less than half a second to almost twenty seconds. As we could not determine the exact moment which was reflected in the response (it could be the beginning of the processing as well as the middle or the end), we decided to keep a fixed 10-second distance between receiving the response and sending a new request. Finally, for this single week, we logged 58,629 states with the average distance between two consecutive measurements equal to $10.32$s. In 76.43\% of cases, the distance was equal to 10 seconds (Slurm returned the response to the previous request almost immediately); in 23.26\% of cases, it was between 11 and 13 seconds.
Higher distances were observed in the remaining 0.31\% of cases. Thus, for the \emph{Slurm-level} perspective, we will always assume that the logged states are equally-distanced in time (ca. 10 seconds) and the average distance will always be provided.

After each experiment, we try to answer an additional question of how efficient Slurm was in covering node availability periods. Thus, we introduce the third perspective --- \emph{Simulation}. Namely, based on the \emph{Slurm-level} data, we perform an a posteriori, clairvoyant simulation that provides us with an upper bound on the potential node-coverage within a given job model: \emph{fib} or \emph{var}. This helps us evaluate the efficiency of the actual scheduling decisions.

\subsection{Job lengths for our experiments}
\label{subsec:lengths}

There is a noticeable amount of time between the start of an HPC-Whisk job on a cluster node and the moment when a booted up OpenWhisk invoker running inside of this HPC-Whisk job registers in the controller as healthy. Our experiments revealed that the median warm-up time of an HPC-Whisk job on \clustername{} was 12.48s with the 95th percentile of 26.50s.
This measured initialization time is used as a parameter during our a posteriori simulations, e.g., in Sect.~\ref{subsec:coverage} when we compare the coverage of the idle periods between the offline simulator and the actual execution.

In the \emph{fib} model, we maintain a queue of preemptible jobs with fixed lengths. We use our simulator to optimize the set of lengths that maximizes the coverage of the idleness periods with healthy OpenWhisk workers. These lengths need to factor in two opposite effects: (1) short jobs will be easy to fit into the idle periods, but having only short jobs will cause frequent, unnecessary warm-ups; (2) long jobs reduce the number of warm-ups, but are harder to fit.

Due to the cluster configuration, we only consider jobs with even lengths, as the backfill scheduler operates on 2-minute slots. Namely, if we used jobs with odd lengths, we would loose one minute of possible computing time, as such jobs would be fitted into even-length windows. The declared run time of a job is between 2 minutes, which is the length of a slot, and 120 minutes, which is backfill's \emph{window} --- the period of time in the future being considered for executing jobs.
Thus, we compare 6 sets of potential job lengths, including those based on the Fibonacci sequence (A1-A3), powers of 2 (set B), and reflecting Slurm's variable job length models for even-length allocation slots (C1-C2).

We test different variants of the Fibonacci sequence, as it models replacing two shorter jobs with a single longer job, each time saving a single warm-up time. 
In contrast, the exponential series (set B) may result in a disproportional increase in the number of allotted jobs if the idleness period is slightly shorter than the length of one of such jobs. For example, if a node is idle for 62 minutes, it would be allocated 5 set-B jobs, while only 2 or 3 jobs from sets A1-A3.

For each of the sets, we performed an offline simulation based on the log data gathered between 21st and 27th Feb. 2022. The simulator greedily fills each period of idleness with the jobs, starting from the longest ones that fit.
For example, when we considered set A1 and node $x$ that was idle for 21 minutes, we allotted it with jobs of 14 and 6 minutes, respectively, and 1 minute was not used. 
Based on the warm-up time analysis, we assume that the first 20 seconds of each job are lost (counted as \emph{warm up} in the results below).

Table~\ref{tab:presim} presents the detailed results of our simulations. For each of the sets of job lengths, we present the total number of jobs that were allotted, the share of idle time in each of the states, and the distribution of the number of ready workers in time. We also present the total share of time where no worker was ready.

Table~\ref{tab:presim} shows that the choice of the set of job lengths has no significant impact on the overall performance, although set A1 presented a slightly higher share of idle time used by ready nodes and thus the higher average number of ready workers when compared to A2 and A3. For this reason, we will use job lengths from the set A1 (for the \emph{fib} model) and C2 (for the \emph{var} model).

\begin{table*}
\def\arraystretch{0.9}
\caption{The simulation of the coverage of idleness periods between 02/21 and 02/27/2022. We assume that an HPC-Whisk worker requires 20 first seconds of the job to warm up. The maximum job length is 120 minutes.}
\label{tab:presim}
\begin{tabularx}{\textwidth}{clccccccccc}
\toprule
&
&
&
\multicolumn{3}{c}{Share of idle time by state} &
\multicolumn{4}{c}{\# of ready workers} & \\

Set &
Job lengths [min] &
\# of jobs &
\emph{warm up} &
\emph{ready} &
not used &
25\%- &
50\%- &
75\%-ile &
Avg  &
Non-availability [\%] \\
\midrule

A1 &
2, 4, 6, 8, 14, 22, 34, 56, 90 & 
10767 & 
3.98\% & 
80.58\% & 
15.44\% & 
2 & 
4 & 
8 & 
7.44 & 
14.82\% \\

A2 &
2, 4, 8, 12, 20, 34, 54, 88 & 
11659 & 
4.31\% & 
80.25\% & 
15.44\% & 
1 & 
4 & 
8 & 
7.41 & 
14.85\% \\

A3 &
2, 4, 6, 10, 16, 26, 42, 68, 110 & 
11258 & 
4.16\% & 
80.40\% & 
15.44\% & 
2 & 
4 & 
8 & 
7.42 & 
14.84\% \\

B &
2, 4, 8, 16, 32, 64 & 
12348 & 
4.56\% & 
80.00\% & 
15.44\% & 
1 & 
4 & 
8 & 
7.38 & 
14.90\% \\

C1 &
2, 4, 6, 8, 10, 12, 14, 16, 18, 20 & 
10651 & 
3.94\% & 
80.63\% & 
15.44\% & 
2 & 
4 & 
8 & 
7.44 & 
14.78\% \\

C2 &
2, 4, 6, \dots, 120 & 
9115 & 
3.37\% & 
81.20\% & 
15.44\% & 
2 & 
4 & 
8 & 
7.49 & 
14.73\% \\

\bottomrule
\end{tabularx}
\end{table*}

\section{Efficiency of the Prototype --- Experiments}
\label{sec:efficiency}

Here we describe the results of running HPC-Whisk on \clustername{}. The goal of these experiments is to estimate the performance of HPC-Whisk as deployed on a production HPC system. We measure the following aspects of performance.

\begin{enumerate}
    \item We measure how efficient HPC-Whisk is in converting idle periods of the production cluster into readiness periods of FaaS workers (Sect.~\ref{subsec:coverage}). This experiment runs for two periods of 24h on a large-scale, production HPC cluster.
    This experiment essentially benchmarks the HPC-Whisk job manager coupled with Slurm.
    \item We measure the FaaS-client perspective on the stability of the constructed FaaS infrastructure. In particular, we quantify how the infrastructure absorbs frequent failures of worker nodes (when e.g. they are preempted by higher-priority HPC jobs) (Sect.~\ref{subsec:responsiveness}). 
    Frequent failures may result in the system that frequently loses or rejects function invocations --- thus making it difficult to use.
    \item We estimate the computational power of a single HPC-Whisk worker by running functions from SeBS~\cite{copik2021sebs}, a standard FaaS benchmark, and comparing the runtimes to AWS Lambda.
    HPC-Whisk runs essentially for free (as otherwise, the nodes would be idle) --- however, we want to check whether there is a performance penalty when switching to a makeshift solution from a specialized, FaaS-optimized infrastructure AWS Lambda offers.
\end{enumerate}

These three aspects show how efficiently HPC-Whisk is in transforming the ``raw'', idle surface of an HPC cluster into processing meaningful FaaS workloads.

\subsection{Method}
As we propose two models of supplying OpenWhisk jobs to Slurm, \emph{var} and \emph{fib}, we performed separate experiments for each of these models. The initial analysis (see Sect.~\ref{sec:introduction}) showed no significant weekly patterns in the number of idle nodes throughout the week, so we decided that each of our experiments would last for 24 hours. It allowed us to reduce the influence of external factors (e.g. time of the day, dynamic structure of the queue of HPC jobs, technical maintenance of selected nodes, etc.) on our results. Therefore, we purposely chose two separate full working days for our experiments.

\subsection{Node coverage with HPC-Whisk jobs}
\label{subsec:coverage}

During the experiment, we repeatedly logged the state of the Slurm nodes based on the technique described in Sect.~\ref{exps:method}. In our initial analysis, all the nodes which were not executing HPC jobs were considered idle. In contrast, during our experiments, available nodes could either be waiting for a job (idle), or executing one of our HPC-Whisk jobs. Thus, we logged the lists of nodes in those two states, separately.

Our HPC-Whisk jobs had the lowest priority tier of 0. Thus, if we supplied no HPC-Whisk jobs, then a node that finally executed such a job would be idle. For this reason, we joined the sets of both idle and HPC-Whisk nodes to create a baseline for our a posteriori analysis.

\subsubsection{The \emph{fib} model}

The experiment for the \emph{fib} model was performed on March 17th, 2022. Based on the analysis from Sect.~\ref{subsec:lengths}, we supplied Slurm with jobs of 9 different lengths, i.e. 2, 4, 6, 8, 14, 22, 34, 56, and 90 minutes. 

After 24 hours, we analyzed the \emph{Slurm-level} logs which consisted of 8,057 measurements with an average distance of 10.72s. First, we joined the sets of idle and HPC-Whisk nodes to analyze the overall ``HPC-idle'' surface of the measured period. We concluded that, on average, $11.85$ nodes were available (with a median of $11$). No node was available for $51$ logged states ($0.6\%$ of the time).

\begin{figure*}[tb]
    \centering
    \subfloat[Number of OpenWhisk worker jobs running on the \clustername{} cluster on 03/17/2022 and number of the remaining idle nodes. From the left, we present three perspectives: a posteriori simulation based on the Slurm-level logs, Slurm-level logs, and the actual number of worker jobs reachable from the OpenWhisk-level in time. The warming-up workers are hardly-visible on the charts as their average number is 0.4 and 0.06, respectively. In first two charts, the left Y-axis corresponds to the number of worker jobs, and the independent right Y-axis to the number of idle jobs.]{{
        \includegraphics[width=0.31\textwidth,clip,valign=t]{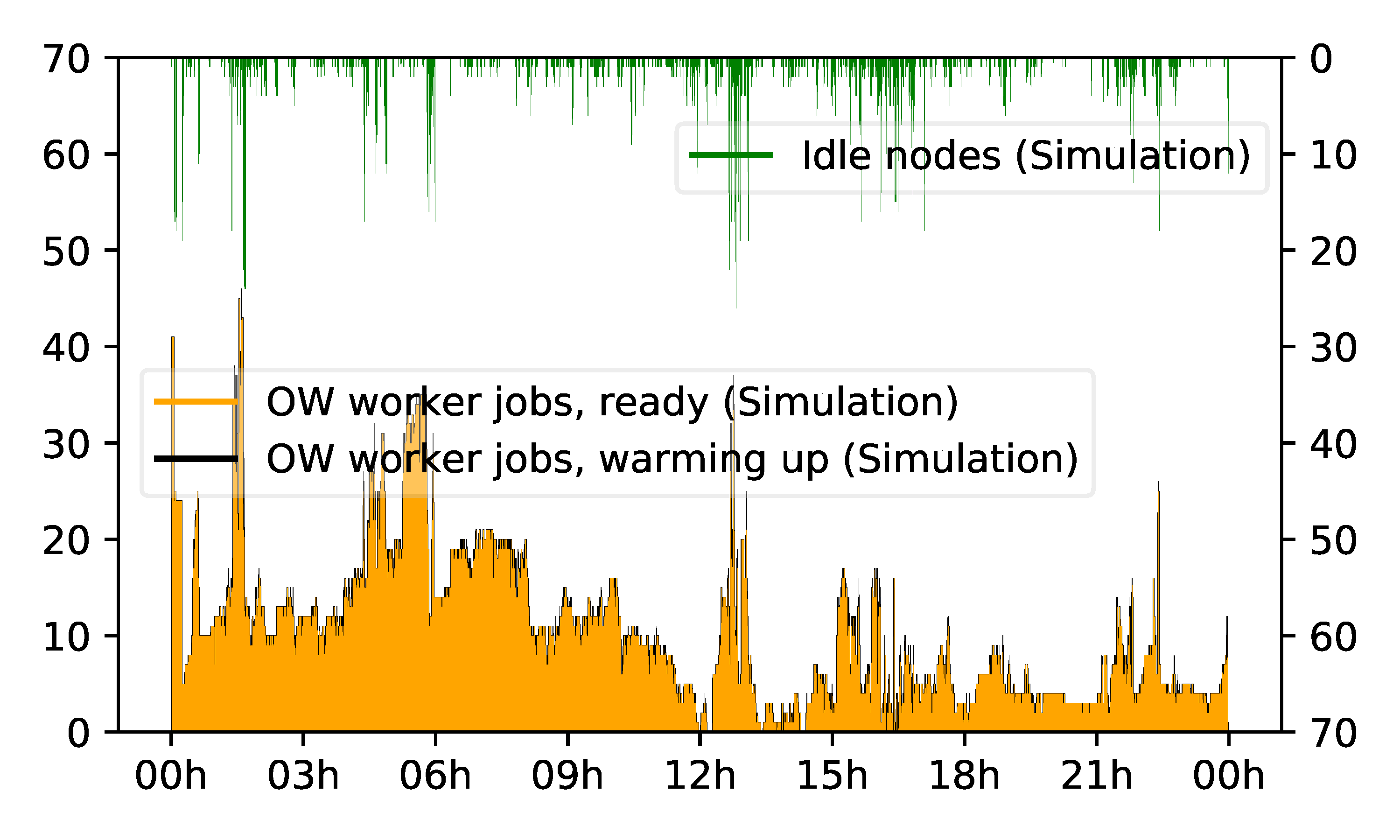}
        \includegraphics[width=0.31\textwidth,clip,valign=t]{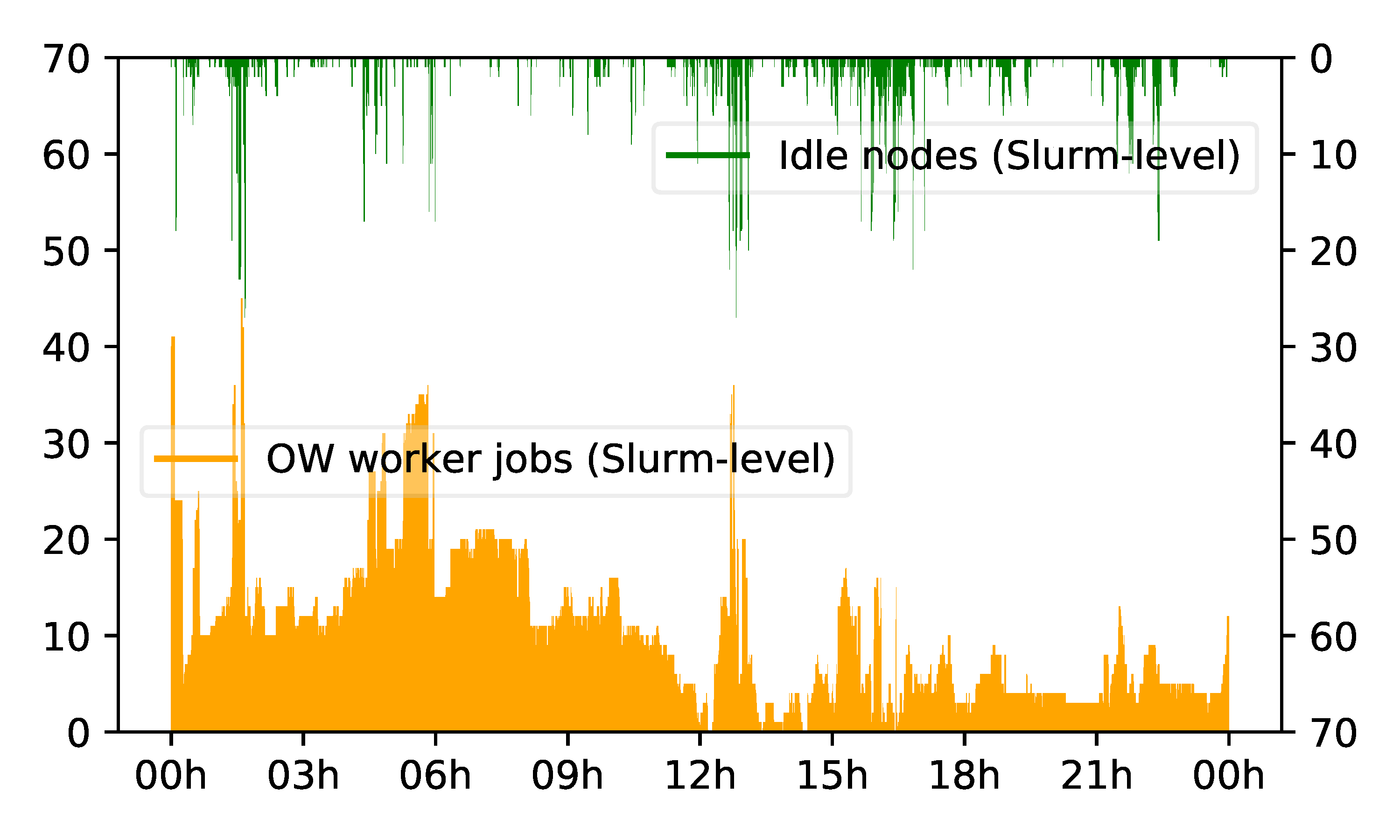}
        \includegraphics[width=0.31\textwidth,clip,valign=t]{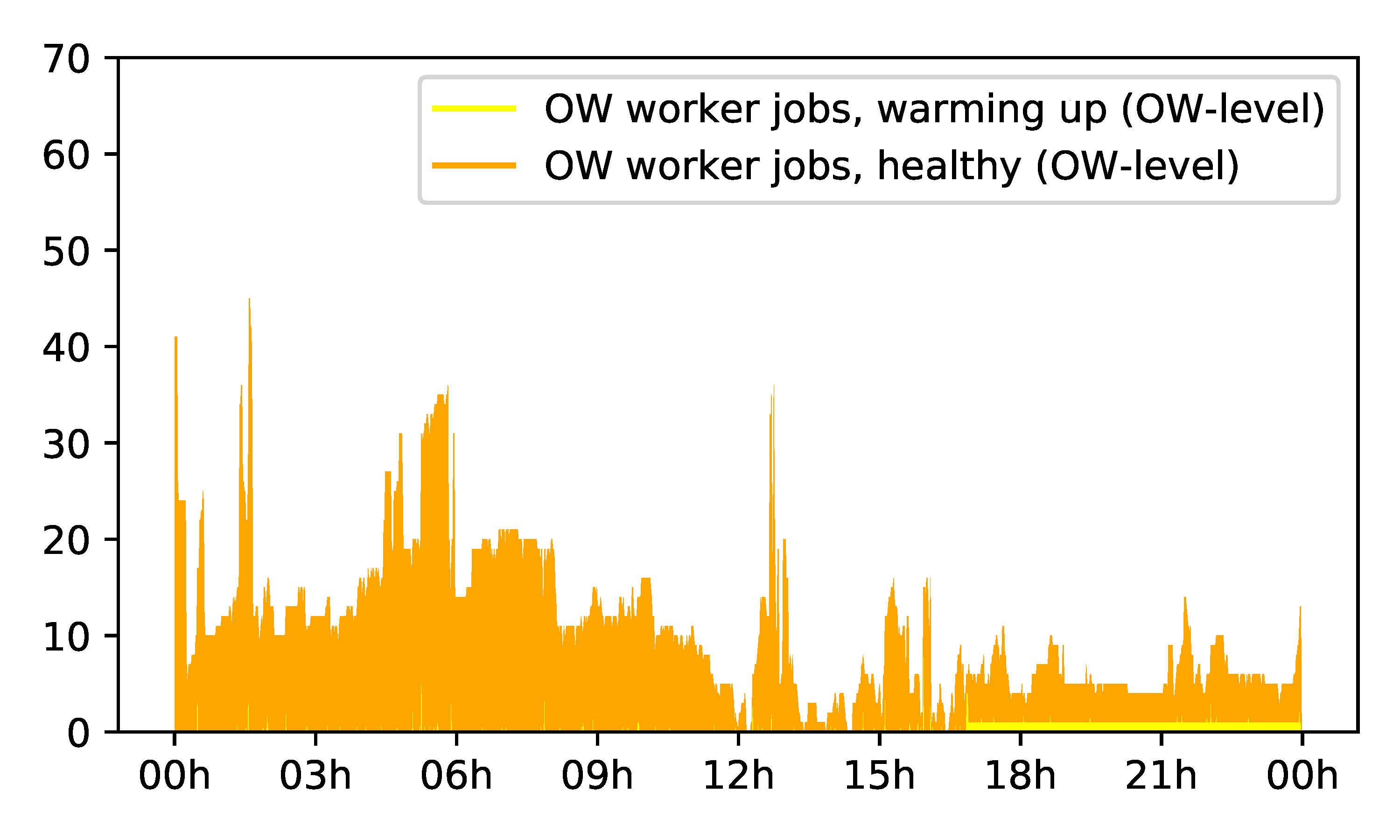}
        \label{exp-1-workers}
    }}%
    
    \subfloat[Number of successful, failed and lost queries over time. Each point shows an aggregate over a minute. A system with a steady load of 10 QPS.]{
        \includegraphics[width=0.43\textwidth,clip,valign=t]{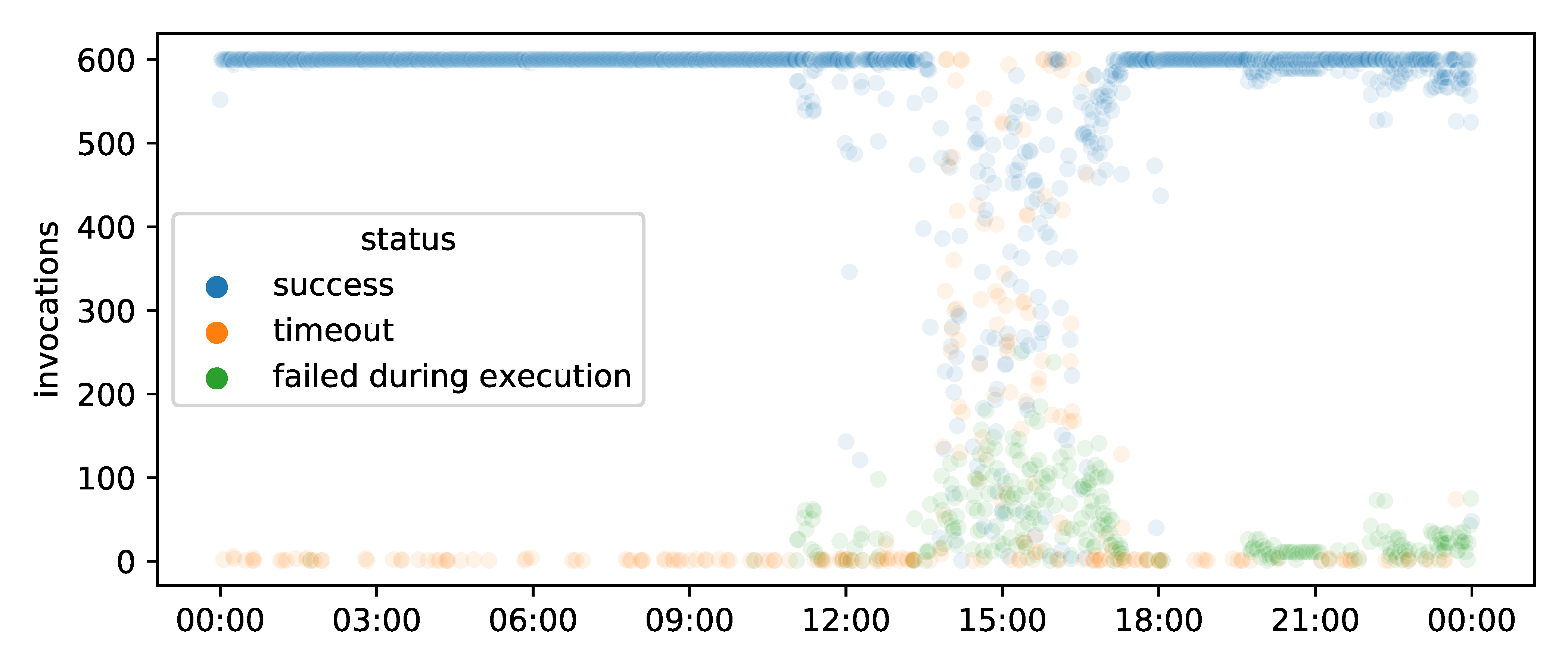}
    }\hspace*{1em}%
    \subfloat[The CDF presents the distribution of idle nodes (green), OpenWhisk nodes (orange) and originally-idle nodes (black) --- Slurm-level analysis.]{{
        \includegraphics[width=0.31\textwidth,clip,valign=t]{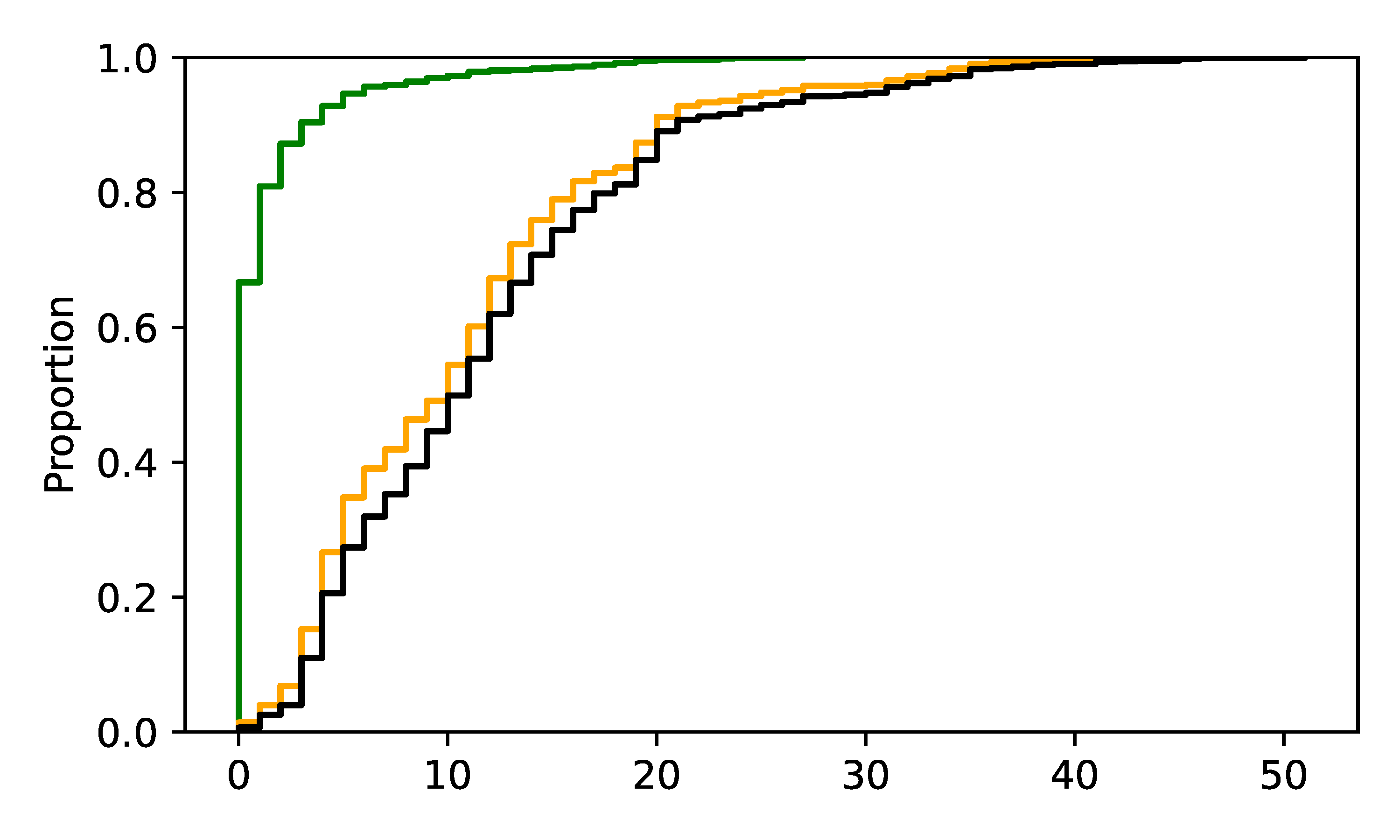}
        \label{exp-1-idle}
    }}%
      
    \caption{Comparison of the results for the experiment based on the \emph{fib} model.}
	\label{fig:all_cmp_fib}
\end{figure*}

Then, we estimated the actual idle node coverage to be $90\%$.
This means that the \clustername{} cluster was executing HPC-Whisk jobs for 90\% of the time that would otherwise be idle.
Our clairvoyant, a posteriori simulation based on the A1 set revealed that the maximum share of availability time that we could utilize for HPC-Whisk jobs with our \emph{fib} job manager can be estimated by $92\%$.
Thus, we conclude that the simulator can be used to estimate the potential of HPC-Whisk on a cluster without the need to configure the software.
Fig.~\ref{exp-1-workers} presents the time-series for all three perspectives, including the actual results gathered on the OpenWhisk-level. In Fig.~\ref{exp-1-idle} presents a CDF of OpenWhisk worker jobs in different states. Tab.~\ref{tab:fib_cmp_numbers} summarizes the results.

The average number of healthy (ready) invokers on the OpenWhisk-level reached 10.39, which is slightly less than the estimated 10.66 (Slurm-level) and simulated 10.59 (Simulation). The median number of ready workers was 9, which is one less than the simulated value. The total time when no invoker was reachable by the OpenWhisk controller was 24 min, and the longest continuous period of unavailability was 7 min (starting at 2:19PM). The OpenWhisk invoker was ready to process incoming function calls for an average of over 23 min (with the median of slightly less than 11 minutes, and the 75th percentile of almost 31 min).

\begin{table}
\def\arraystretch{0.9}
    \centering
    \caption{The comparison of the results obtained on our setup, for \emph{fib} job manager, on 03/17/2022.}
    \label{tab:fib_cmp_numbers}
    \begin{tabularx}{\columnwidth}{llcccc}
         \toprule
         & & \multicolumn{2}{c}{\# of workers} & \multicolumn{2}{c}{Share of idle time} \\
         & & 25-50-75p & avg & used & not used \\
         \midrule 
         Simulation & \emph{warm up} & 0-0-0 & 0.31 & 2.61\% & \multirow{2}{1.5cm}{\centering 8.05\%} \\
         & \emph{ready} & 4-10-14 & 10.59 & 98.34\% & \\
         \midrule
         Slurm-level & \emph{all states} & 4-10-14 & 10.66 & 89.97\% & 10.03\% \\
         \midrule
         OW-level & \emph{warm up} & 0-0-1 & 0.40 & & \\
         & \emph{healthy} & 4-9-14 & 10.39 & & \\
         & \emph{irresp.} & 0-0-0 & 0.06 & & \\
         \bottomrule
    \end{tabularx}
\end{table}

\subsubsection{The \emph{var} model}

The experiment for the \emph{var} model was performed on March 21st, 2022. We supplied Slurm with jobs of flexible lengths between 2 and 120 minutes. The exact execution time was determined by Slurm based on its inner scheduling policies.
After 24 hours, we analyzed the logs which consisted of 8,091 measurements with an average distance of $10.68$s. On average, $7.38$ workers were available (with a median of $6$). No worker was available in 764 logged states ($9.44\%$). We observed a significant difference in the available surface between  the \emph{fib} and \emph{var} experiments.

\begin{figure*}[tb]
    \centering
    \subfloat[Number of OpenWhisk worker jobs running on the \clustername{} cluster on 03/21/2022. From the left, we present three perspectives: a posteriori simulation based on the Slurm-level logs, Slurm-level logs, and the actual number of worker jobs reachable from the OpenWhisk-level in time. The warming-up workers are hardly-visible on the charts as their average number is 0.07 and 0.15, respectively. In first two charts, the left Y-axis corresponds to the number of worker jobs, and the independent right Y-axis to the number of idle jobs.]{{
        \includegraphics[width=0.31\textwidth,clip,valign=t]{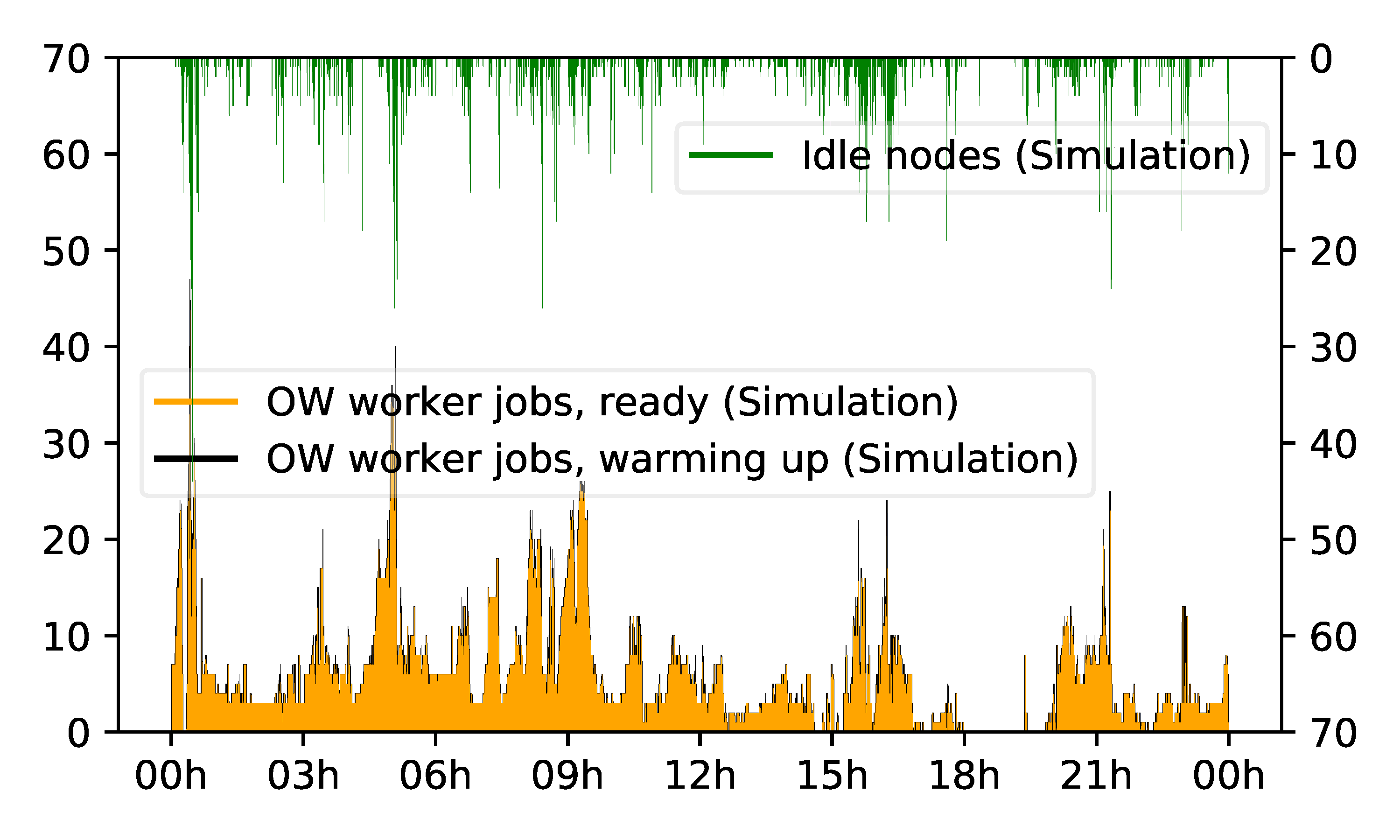}
        \includegraphics[width=0.31\textwidth,clip,valign=t]{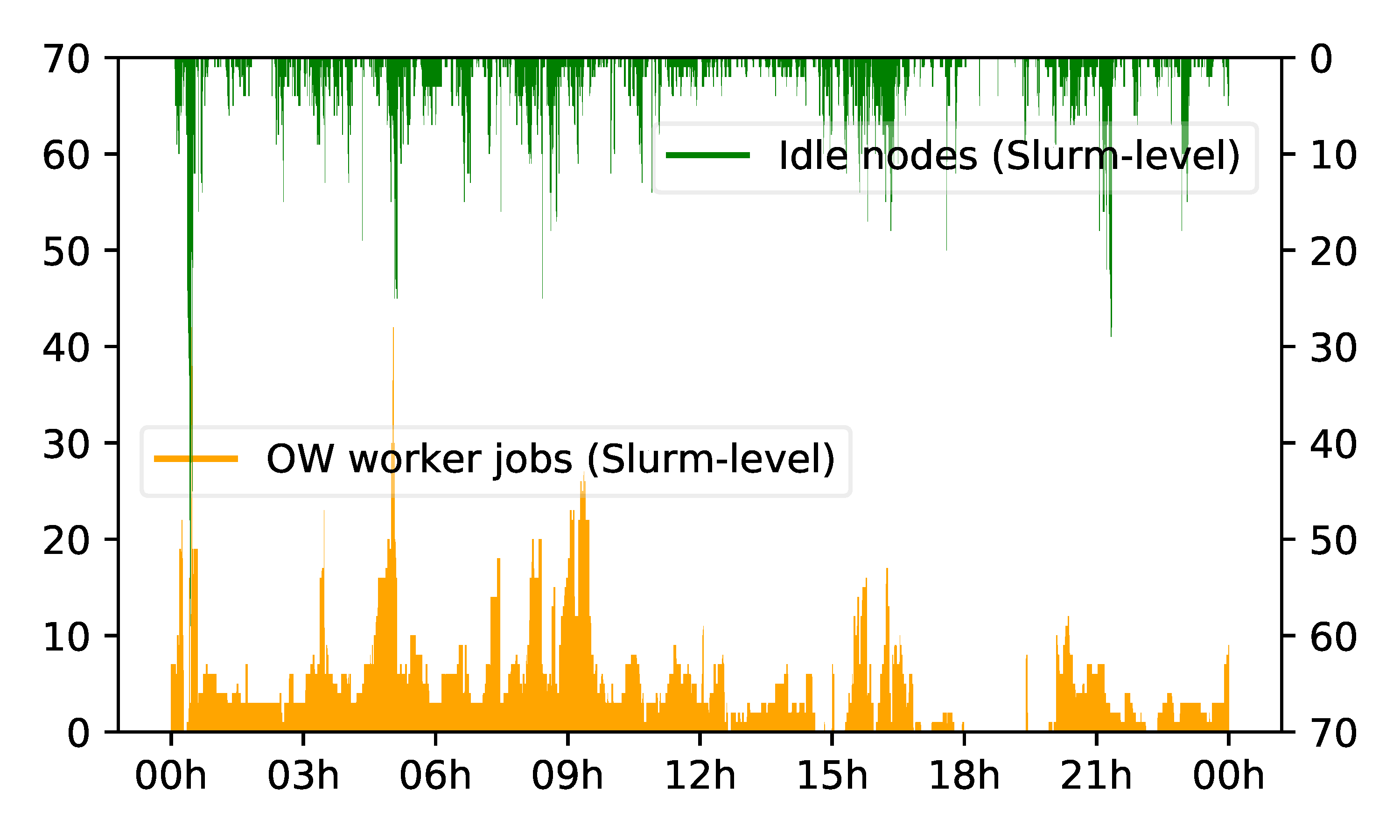}
        \includegraphics[width=0.31\textwidth,clip,valign=t]{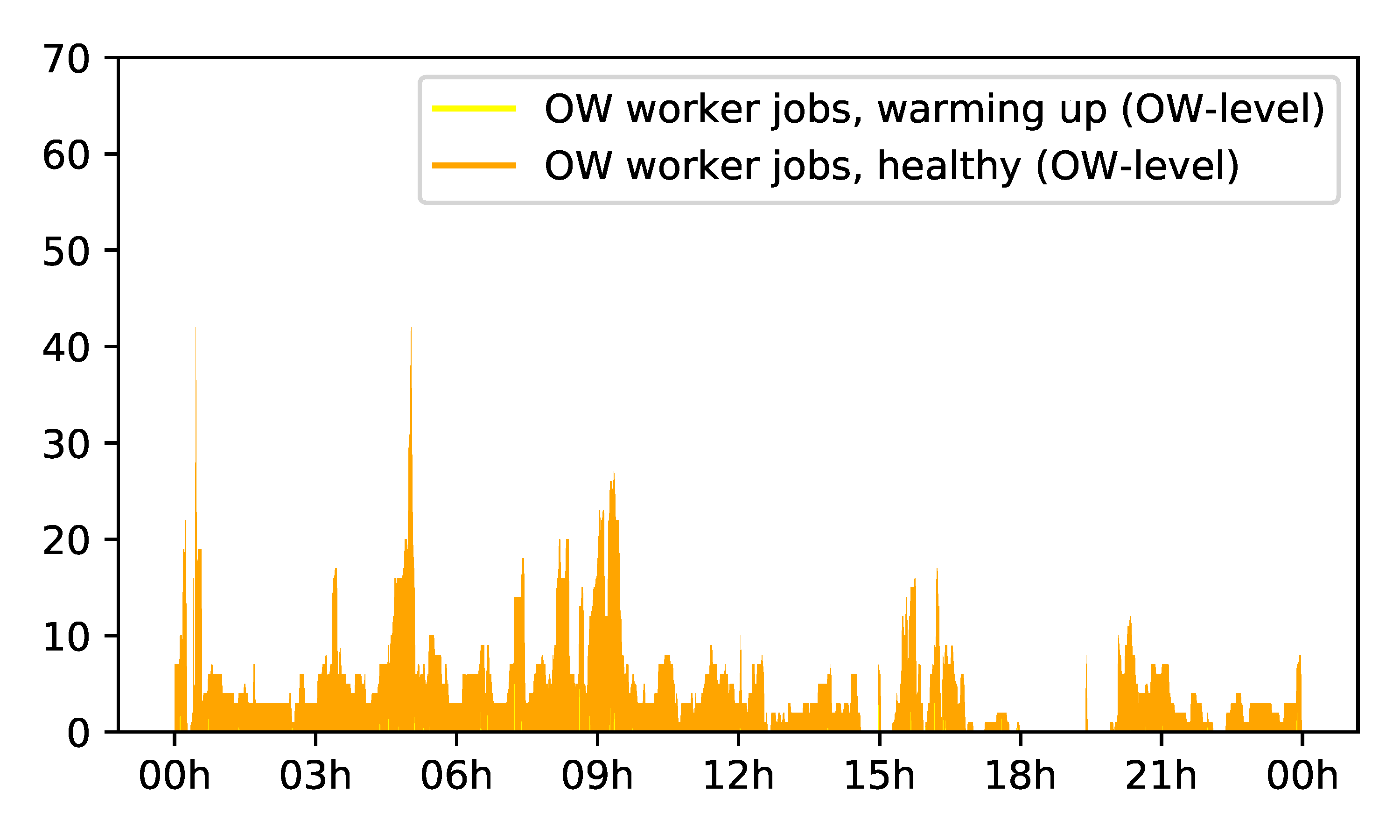}
        \label{exp-2-workers}
    }}%
    
    \subfloat[Number of successful, failed and lost queries over time. Each point shows an aggregate over a minute. A system with a steady load of 10 QPS except when there are no OpenWhisk workers (at around 18:00).]{
        \includegraphics[width=0.43\textwidth,clip,valign=t]{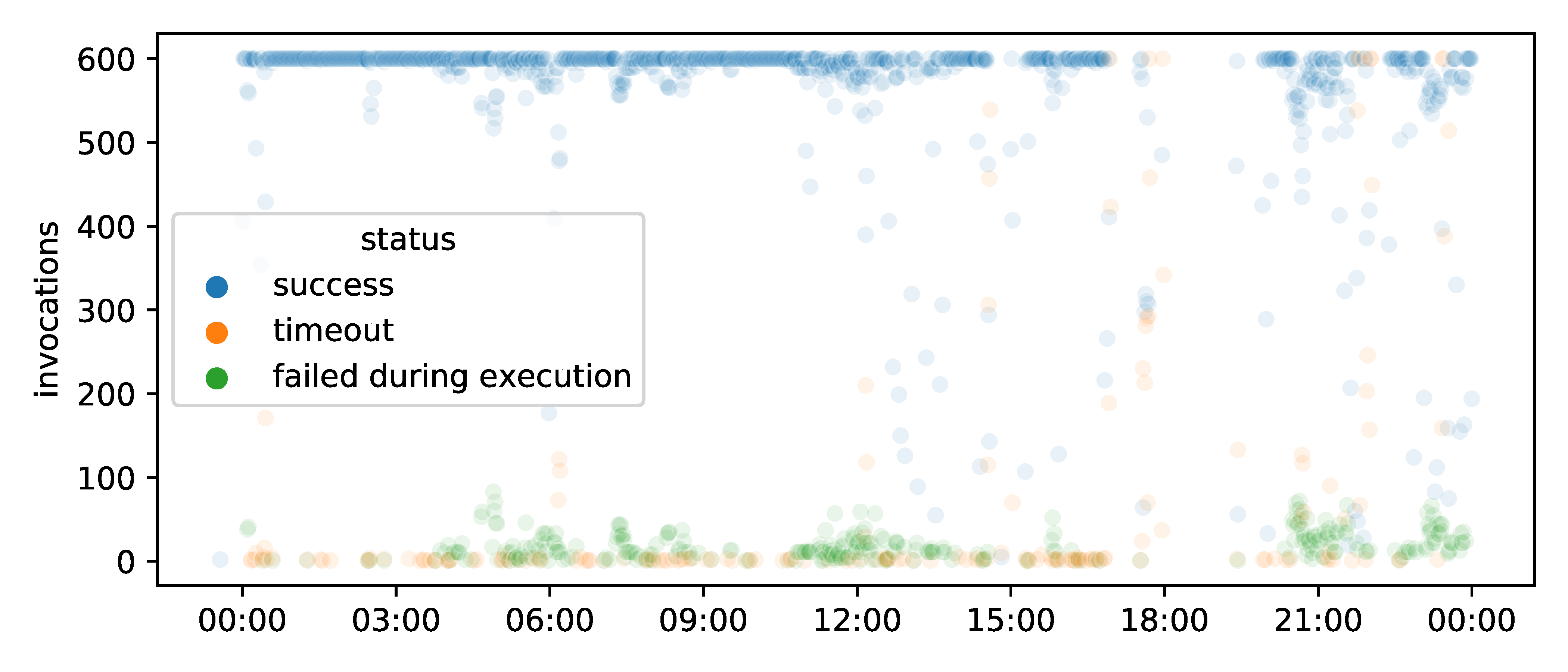}
    }\hspace*{1em}%
    \subfloat[The CDF presents the distribution of idle nodes (green), OpenWhisk nodes (orange) and originally-idle nodes (black) --- Slurm-level analysis.]{{
        \includegraphics[width=0.31\textwidth,clip,valign=t]{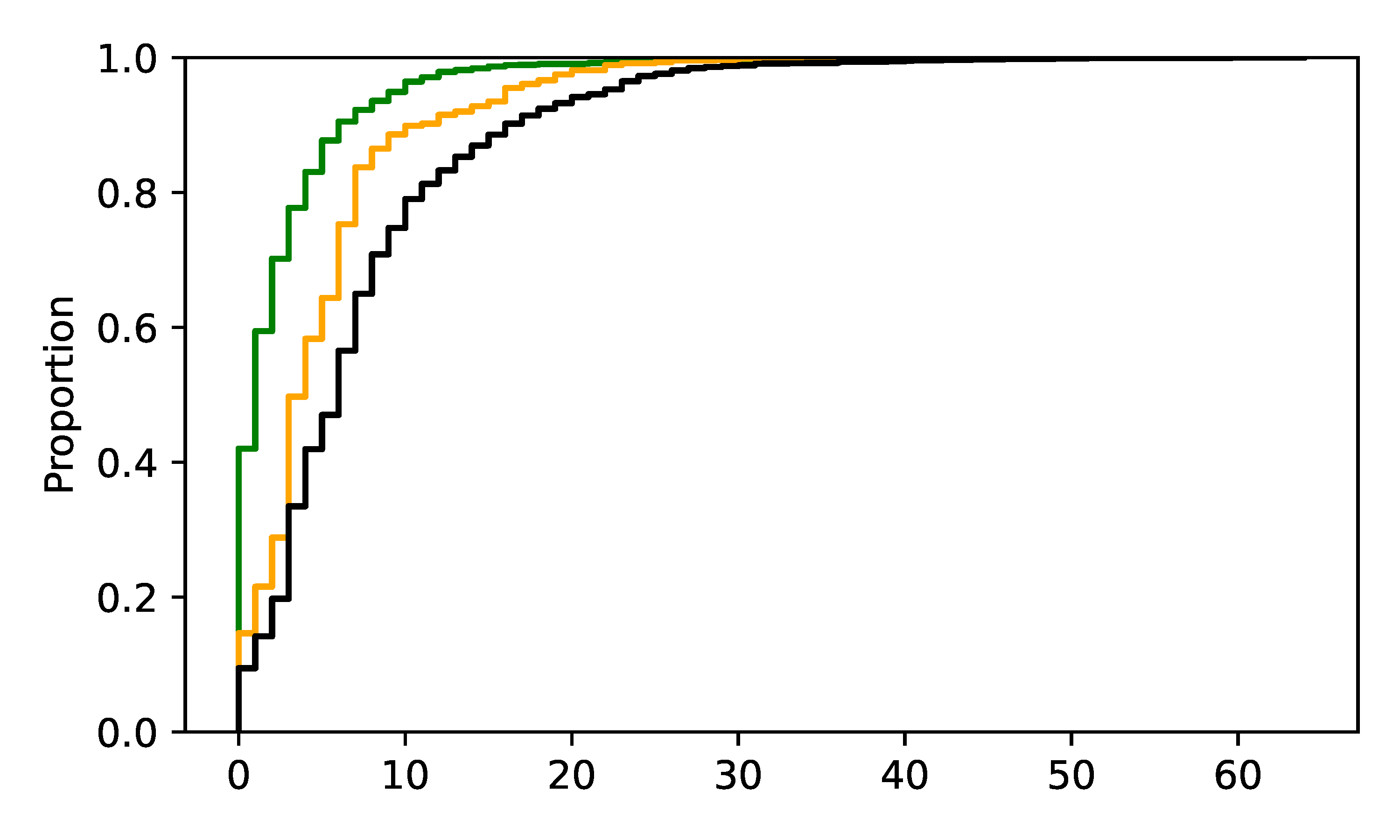}
        \label{exp-2-idle}
    }}%
      
    \caption{Comparison of the results for the experiment based on the \emph{var} model.}
	\label{fig:all_cmp_var}
\end{figure*}

We estimated the actual idle node coverage to be $68\%$. Our a posteriori simulation based on the C2 set revealed that the maximum share of availability time that we could utilize for HPC-Whisk jobs with our \emph{var} job manager can be estimated by 84\%.
The reason for such a difference between the simulated and the Slurm-level estimated coverage might lie in the Slurm-level procedures of scheduling variable-length jobs. Such a procedure involves scheduling a job with the minimum time requirement and then extending it until the time limit is reached or available resources are exhausted. Due to the complexity of this procedure, the scheduler may not be able to process the queue before the environment changes, which directly impacts the scheduling effectiveness.
In Fig.~\ref{exp-2-workers} we present the time-series results for all three perspectives. In Fig.~\ref{exp-2-idle}, we present a CDF of OpenWhisk worker jobs in different states. In Tab.~\ref{tab:var_cmp_numbers}, we present aggregate statistics.

We observe that the average number of healthy invokers on the OpenWhisk-level reached 4.96, which is slightly less than the estimated 5.03 (Slurm-level) and significantly less than the simulated 5.97 (Simulation). The median number of ready workers was 3, which is also one less than the simulated value. The total time when no invoker was reachable by the OpenWhisk controller was 218 min, and the longest continuous period of such property was 85 min (starting at 5:59PM). The OpenWhisk job was ready to process incoming function calls for an average of over 14 min (with the median of slightly less than 7 min, and the 75th percentile of 14.5 min).

\begin{table}
\def\arraystretch{0.9}
    \centering
    \caption{The comparison of the results obtained on our setup, for \emph{var} job manager, on 03/21/2022.}
    \label{tab:var_cmp_numbers}
    \begin{tabularx}{\columnwidth}{llcccc}
         \toprule
         & & \multicolumn{2}{c}{\# of workers} & \multicolumn{2}{c}{Share of idle time} \\
         & & 25-50-75p & avg & used & not used \\
         \midrule 
         Simulation & \emph{warm up} & 0-0-0 & 0.23 & 3.18\% & \multirow{2}{1.5cm}{\centering 15.87\%} \\
         & \emph{ready} & 3-4-8 & 5.97 & 80.95\% & \\
         \midrule
         Slurm-level & \emph{all states} & 2-4-5 & 5.03 & 68.20\% & 31.80\% \\
         \midrule
         OW-level & \emph{warm up} & 0-0-0 & 0.07 & & \\
         & \emph{healthy} & 2-3-6 & 4.96 & & \\
         & \emph{irresp.} & 0-0-0 & 0.15 & & \\
         \bottomrule
    \end{tabularx}
\end{table}

\subsection{Responsiveness of our setup}
\label{subsec:responsiveness}

As the controller determines the target invoker based on the name of the called function, we deployed 100 identical functions with different names to always utilize as many warmed-up invokers as possible. The functions were based on the \texttt{sleep} system call, as we wanted to measure the responsiveness of the system, not its performance (which we measured separately in Sect.~\ref{subsec:perf-node}). Each of the functions required $10$ ms to execute. During the experiments, we continuously called these functions from outside of the cluster with a constant rate of 10 calls per second (864,000 requests in total) which corresponds to a 10\% load of a single node. We used Gatling \cite{Gatling} to both make calls and collect the responses.

On March 17th (the \emph{fib} experiment), 95.29\% of requests were invoked by the OpenWhisk controller (and 4.71\% resulted in error 503). For these calls, Alg.~\ref{alg:wrapper} would pass the invocations to a commercial FaaS service. In total, 95.19\% of all successfully invoked requests ended with success. Slightly over 3\% of all such requests were not finished (timeout), and the remaining 1.65\% failed during execution. On this  day, between 14:30 and 17:00, the running invokers achieved the upper limit of concurrently running container processes which resulted in an increased number of failed invocations. The median response time of successfully-executed calls on the Gatling level was 865ms. The difference between 10ms of internal execution time and roughly 0.8s of the response time is similar to the results obtained for similarly-short functions from SeBS on AWS Lambda (see \cite{copik2021sebs}, Fig. 3).

On March 21st (the \emph{var} experiment), 78.28\% of requests were invoked by the OpenWhisk controller (and 21.72\% resulted in 503 error). In total, 96.99\% of all successfully invoked requests ended with success (with the median response time on the Gatling-level equal to 1,227ms due to the insufficient number of active invokers). Slightly over 2\% of all such requests were not finished (timeout), and the remaining 0.9\% failed during execution. 

\subsection{Performance of single invocations}
\label{subsec:perf-node}

\begin{figure}
    \centering
    \includegraphics[width=0.85\columnwidth]{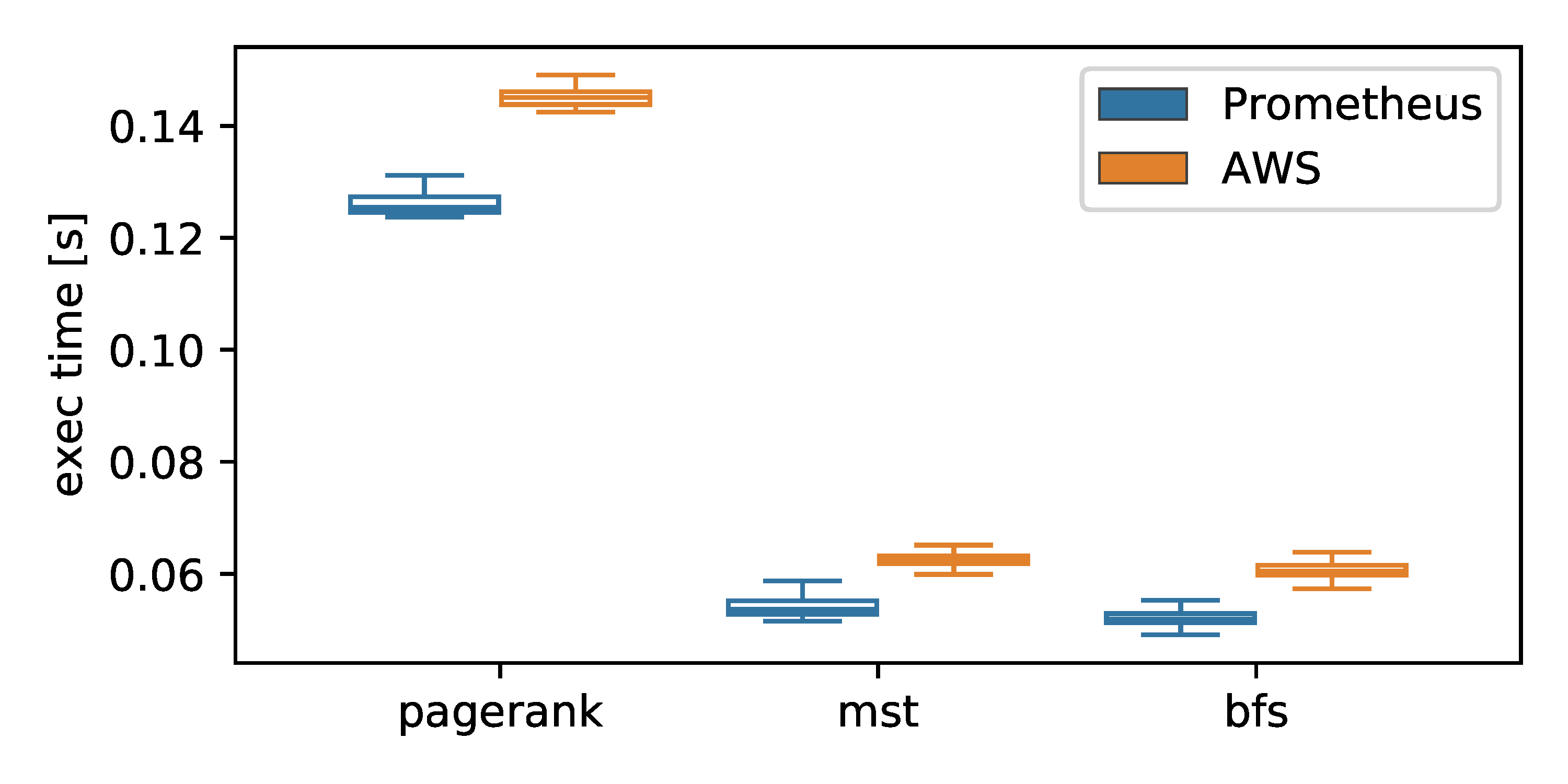}
    \caption{Performance comparison of a single \clustername{} node and AWS Lambda with 2048MB RAM}
    \label{fig:single-node}
\end{figure}

In a final series of experiments, we study how the efficiency of \clustername{} nodes compares with standard FaaS commercial offerings. We focus on compute-intensive functions, as we want to compare the efficiency of the nodes, and not the network connection quality or local storage speed --- scientific applications of FaaS will most probably be run locally on a cluster.
We run the same set of three compute-intensive benchmarks (\texttt{bfs}, \texttt{mst} and \texttt{pagerank}) from the SeBS suite on \clustername{} and on AWS Lambda.
Those chosen functions are classified as compute-intensive, as they do not use any storage or network resources.
For each function, we perform 200 invocations to focus on ``warm'' performance, i.e. without the overhead of function's initial execution.
As we do not want to benchmark the network connections, we only report the internal execution time of each invocation.
As observed in~\cite{copik2021sebs}, AWS Lambda performance depends on the amount of declared memory, as it correlates with the share of CPU capacity available for this function --- we report the results only on the fastest configuration (with 2GB RAM). 
Fig.~\ref{fig:single-node} shows the results.
The results for AWS Lambda essentially reproduce the results from~\cite{copik2021sebs}, thus validating our test configuration.
The results for \clustername{} show a roughly 15\% performance improvement consistent across all the functions.
This slight performance gain can be explained by the HPC cluster's machines hardware architecture optimized for compute-intensive workloads.
This result shows that, at least for compute-intensive functions, an HPC cluster has comparable performance to a high-end commercial cloud.

\section{Related work}\label{sec:related-work}

This work is related to three main areas: serverless computing in the HPC context, using idle HPC resources for other non-HPC tasks, and pilot job scheduling on HPC clusters.

Regarding the usage of serverless computing for scientific computing which typically requires HPC, there are examples of projects which try to re-purpose existing FaaS solutions for scientific workloads. These examples are mainly focusing on high-throughput independent tasks, such as in PyWren~\cite{jonas_occupy_2017} or on workflows (graphs of tasks) as in HyperFlow~\cite{malawski_serverless_2020}. There are also successful attempts to migrate typical tightly-coupled computations such as dense linear algebra to serverless model~\cite{shankar_serverless_2020}. On the other hand, frameworks such as FuncX~\cite{chard_funcx_2020} allow running FaaS tasks on supercomputers, demonstrating scalability to thousands of concurrent executions and the capability to combine HPC and cloud resources. Our work is different from FuncX in that FuncX uses regular scheduling facilities of HPC clusters to spawn FaaS infrastructure on top of HPC, while we focus on using idle resources from the cluster to build a running FaaS system. 

Using idle HPC resources for other tasks has been studied in the context of Big Data  and other non-HPC workloads. For example, combining the OAR job management system on the HPC cluster with the Hadoop YARN scheduler for analytics jobs allows the authors of~\cite{mercier_big_2017} to reach a full cluster utilization, with an observed penalty of 17\% on the mean waiting time for HPC jobs. A different approach~\cite{souza_hybrid_2019} proposes a hybrid scheduling by combining Slurm and Mesos and sharing computing nodes to allow running HPC and analytics jobs concurrently on the same resources. Such an approach leads to a decrease in average waiting time for both types of jobs, at the cost of increasing the average walltime of the jobs due to their interference with the shared resources. Yet another example, CiGri~\cite{guilloteau_controlling_2021} focuses on Bag-of-Tasks applications which are submitted with the lowest priority to the OAR scheduler in order not to disturb the priority users and not to overload the file server. Our work is distinguished by the feature that we are using idle resources for FaaS jobs without any penalty for regular jobs managed by Slurm on a production HPC cluster.

Our work can be compared to cycle scavenging approaches (e.g., SETI@Home) or, more broadly, desktop grids, which use environments such as BOINC~\cite{seti}. While they solve similar problems with volatile resources and fault tolerance, our focus is on providing a standard FaaS interface and using idle resources not from volunteers but from a production HPC machine.

Our work builds on the concept of pilot jobs~\cite{turilli_comprehensive_2018} known from cluster and grid computing and implemented e.g. in HTCondor~\cite{thain_distributed_2005} and other systems~\cite{glideinwms,saga}. Some solutions like Falkon~\cite{falkon} focus on many fine-grained tasks. Our system can be seen as a pilot-job framework for OpenWhisk on top of Slurm. What is unique in our approach is that although we use standard Slurm  mechanisms, we restrict ourselves to low-priority, preemptible jobs which do not defer any other jobs. Moreover, we provide a standard OpenWhisk interface.

Various backfilling strategies have been developed over the years~\cite{backfill-feitelson,estimation-backfill,characterization-backfill} and implemented in Slurm. Our production cluster already runs a highly optimized aggressive backfill, so a notable achievement of our method is that it is able to find and effectively use the idle nodes which Slurm cannot utilize.

To our best knowledge, there are no other attempts to use idle HPC resources to run FaaS jobs with no penalty for the standard production workload, making our work a unique approach to exploit the potential of unused HPC nodes for the increasingly important serverless workloads in a non-invasive way, by combining standard OpenWhisk and Slurm.

\section{Conclusions}

HPC-Whisk efficiently hosts a convenient, performant FaaS cloud over transient idle nodes of HPC clusters. We analyze logs of a highly-utilized, large-scale HPC cluster with a queuing system tuned over the years.
With long queues and backfilling, the cluster's utilization exceeds 99\%; yet, over a week, the remaining 1\% translates to an idle surface of 37,000 core-hours.
The principal challenge is that this surface is distributed over many 
chunks: a median idle period lasts for just 2 minutes.

HPC-Whisk has three design goals: (1) be minimally invasive on the HPC infrastructure; (2) be convenient for the FaaS users; and (3) be efficient. 
To reduce our footprint on the HPC system software, and not to introduce new, potentially buggy components we rely on pilot jobs we submit to a standard queuing system. These pilot jobs have a priority lower than any prime HPC job; and they are preemptible --- can be terminated while running, if the resources are requested by a newly-submitted HPC job.
From the perspective of FaaS users, HPC-Whisk provides a familiar environment, as it builds upon OpenWhisk.
We keep the same APIs and usage scenarios for the end-users and the function developers.

We optimized efficiency at different layers of the infrastructure.
Our pilot jobs are sized with two strategies: \emph{fib} using jobs with increasing lengths; and \emph{var} relying on the cluster scheduler.
We use simulations to optimize job lengths in the \emph{fib} strategy (Table~\ref{tab:presim}). We then confirm on our production system that \emph{fib} covers almost 90\% of idle surface (Table~\ref{tab:fib_cmp_numbers}). Surprisingly, \emph{var} covers only 68\% (due to scheduler limitations, Table \ref{tab:var_cmp_numbers}).

When using these idle slots, the primary systems challenge is their transient nature: in our experiments, the median duration of a ready (serving) worker exceeded 7 minutes. 
This required us to implement a number of modifications in OpenWhisk to make it more robust.
Our system quickly initializes OpenWhisk workers, as healthy (ready to serve) workers cover over 95\% of the surface of pilot jobs, for both the \emph{fib} and \emph{var} models.
Our new hand-off process for a soon-departing worker results in the successful completion of 95\%-97\% accepted invocations.

Finally, on a node level, we verify that an HPC cluster is a computationally-efficient alternative to commercial FaaS platforms: all three computationally-intensive functions from SeBS benchmark complete 15\% faster than on AWS Lambda.

For some scientific applications, FaaS might be a convenient alternative to monolithic MPI or custom bag-of-tasks solutions: FaaS interfaces are standard and widely used, making development potentially easier; and OpenWhisk middleware is well-established and stable. Our work shows that HPC centers can efficiently execute FaaS workloads in addition to, and without an impact on, their existing workloads.

There are several known limitations of our work, which could be addressed in the future. For example, it would be interesting to benchmark our system using a representative scientific FaaS workload to demonstrate how we can use the unused resources for a real application. Although we did our best to select the representative periods of time for our experiments, it would be interesting to evaluate and characterize the quantity of unused resources in longer periods of time, to identify the potential patterns in the workload which could be of value for the HPC-Whisk job manager. In general, a more in-depth study of typical HPC and FaaS workloads could be pursued, which could lead to optimization of both resource utilization and user experience of both models. 

\section*{Acknowledgments}

The authors thank Damian Sosnowski for his work on implementing the support for Singularity containers in OpenWhisk.

The authors thank Arif Merchant (Google) for his remarks on the early version of the manuscript; and the anonymous reviewers for their help in shaping the paper.

This research was supported by PL-Grid Infrastructure; and by a Polish National Science Center grant Opus (UMO-2017/25/B/ST6/00116). Computations have been performed at Academic Computer Centre Cyfronet AGH, Krakow.

This work is supported by the European Union’s Horizon 2020 research and innovation programme under grant agreement no.
857533 (Sano) and the International Research Agendas programme of the Foundation for Polish Science, co-financed by the
European Union under the European Regional Development Fund.

\balance

\bibliographystyle{IEEEtran}
% \IEEEtriggeratref{12}
\bibliography{main}

\end{document}